\documentstyle[prd,aps]{revtex}

\newcommand{\be}{\begin{equation}}
\newcommand{\ee}{\end{equation}}
\newcommand{\ba}{\begin{eqnarray}}
\newcommand{\ea}{\end{eqnarray}}
\newcommand{\bi}{\begin{itemize}}
\newcommand{\ei}{\end{itemize}}

\newcommand{\nr}[1]{(\ref{#1})}

\newcommand{\RR}{{\rm I\kern -.2em  R}}
\newcommand{\eq}{Eq.~}
\newcommand{\eqs}{Eqs.~}

\begin{document}

\title{LOCAL EQUILIBRIUM OF THE QUARK-GLUON PLASMA}

\author{Cristina Manuel\footnote{Electronic address:
{\tt cristina.manuel@ific.uv.es}}}

\address{\it Instituto de F\'{\i}sica Corpuscular \\
Universitat de Val\`encia-C.S.I.C.\\
Edificio de Institutos de Paterna, Apt 2085 \\
46071 Val\`encia, Spain}

\author{Stanis\l aw Mr\' owczy\' nski\footnote{Electronic address:
{\tt mrow@fuw.edu.pl}}}

\address{\it So\l tan Institute for Nuclear Studies \\
ul. Ho\.za 69, PL - 00-681 Warsaw, Poland \\
and Institute of Physics, \'Swi\c etokrzyska Academy \\
ul. \'Swi\c etokrzyska 15, PL - 25-406 Kielce, Poland}


\maketitle

\begin{abstract}

Within kinetic theory, we look for the local equilibrium configurations 
of a quark-gluon plasma by maximizing the local entropy. We use the 
well-established transport equations in the Vlasov limit,  
supplemented with the Waldmann-Snider collision terms. Two different 
classes of local equilibrium solutions are found. The first one 
corresponds to the configurations that comply with the so-called 
collisional invariants. The second one is given by the distribution 
functions that cancel the collision terms, representing the most probable 
binary interactions with  soft gluon exchange in the $t$-channel. The 
two sets of  solutions agree with each other if we go beyond these 
dominant processes and take into account subleading quark-antiquark 
annihilation/creation and gluon number non-conserving processes.
The local equilibrium state appears to be colorful, as the color 
charges are not locally neutralized. Properties of such an equilibrium 
state are analyzed. In particular, the related hydrodynamic equations 
of a colorful fluid are derived. Possible neutralization processes 
are also briefly discussed.

\end{abstract}

\pacs{PACS: 12.38.Mh, 05.20.Dd, 11.10.Wx}



\section{Introduction}


In the course of equilibration  a many-body system first reaches a local
equilibrium and then it evolves hydrodynamically, usually at a much
slower rate, towards global equilibrium. The distribution function of
local equilibrium is typically of the form of global equilibrium, but
its parameters - temperature, hydrodynamic velocity, chemical potentials
- are space-time dependent. However, the local equilibrium can also
qualitatively differ from the global one. For example, the electron-ion
plasma, which is homogeneously neutral in global equilibrium, can be
locally charged before the global equilibrium is reached, see {\it e.g.}
\cite{Kra73}. Thus,  parameters that are irrelevant for  global
equilibrium  might be needed to describe  local equilibrium. While the
state of global equilibrium is unique, the local equilibrium evolves and
even its qualitative features can change in time. The processes of charge
neutralization are, for example, very fast in the electron-ion plasma.
Therefore, the system is locally neutral after a short time but the
electric currents survive for much longer. Thus, we deal with various
local equilibrium states, depending on the time scale of interest. The
form of local equilibrium is an important characteristics of a system.
Knowing the respective distribution function, one can formulate
a hydrodynamic description of the system. Let us again refer to the
case of the electron-ion plasma. The fact mentioned above that the plasma
is neutralized fast but the currents flow for a longer time justifies
the magneto-hydrodynamics with no electric fields.

The aim of this paper is to discuss local equilibrium of the quark-gluon
plasma. While the global equilibrium features of the system have been studied
in detail, see {\it e.g.} the review \cite{Blaizot:2001nr}, not much is known
about its local equilibrium. Although the problem was formulated long ago
\cite{Heinz:yq,Dyrek:1986vv,Mrowczynski:1987ch,Mrowczynski:1989bv},
the key questions remain unanswered. In particular, the scenario of
equilibration of color degrees of freedom is far not established. It is
unclear whether the regime analogous to magneto-hydrodynamics in
the electron-ion plasma occurs in the quark-gluon plasma. However,
the Yang-Mills magneto-hydrodynamics has been already considered
\cite{Heinz:yq,Mrowczynski:1987ch,Mrowczynski:1989bv,Holm:hg,Holm:yh,Jackiw:2000cd,Bistrovic:2002jx}.

We intend to address these issues which are now of particular interest 
because of the large scale experimental program at the Relativistic Heavy-Ion
Collider (RHIC) in Brookhaven National Laboratory, where high-energy 
nucleus-nucleus interactions are studied, see {\it e.g.} \cite{QM01}. At the 
early stage of such a collision, when the energy density is sufficiently high, 
the generation of the quark-gluon plasma is expected. The most spectacular 
experimental result obtained by now at RHIC is presumably an observation 
of a large magnitude of the so-called elliptic flow \cite{Ackermann:2000tr}.
The phenomenon, which is just sensitive to the collision early stage, is
naturally explained within  hydrodynamics as a result of large density
gradients \cite{Ollitrault:bk}. Since the hydrodynamic description is 
applicable for a system in local thermodynamic equilibrium, the large 
elliptic flow suggests a surprisingly short, below 1 ${\rm fm}/c$ 
\cite{Heinz:2001xi}, equilibration time. Other characteristics of 
relativistic heavy-ion collisions are also consistent with a model
assuming equilibrium state of strongly interacting matter produced
in the collisions, see {\it e.g.} \cite{Broniowski:2002nf}. Thus, 
understanding of the equilibration mechanism of the quark-gluon plasma 
is a key problem for RHIC physics. 

The question of local equilibrium is related to a serious difficulty of 
the transport theory of the quark-gluon plasma. The local equilibrium is defined
as a state which maximizes the local entropy. However, the entropy production
occurs not due to the Vlasov evolution, which is rather well understood
\cite{Blaizot:2001nr,Elze:1989un}, but this is a dissipative phenomenon 
caused by the particle collisions. Thus, the collision terms of the transport
equations are needed to discuss the local equilibrium. However, a derivation 
of these terms has occurred to be a very complex task and only the special 
case of quasi-equilibrium plasma has been seriously examined 
\cite{Selikhov:br,Bodeker:1998hm,Arnold:1998cy,Blaizot:1999xk,Litim:1999id}. 
Fortunately, the structure of the collision terms can be guessed referring to
the analogies between the spin and color systems. And this is not only
a superficial similarity of degrees of freedom governed by the SU(2)
and SU(3) group, respectively. The relationship appears to be much deeper.
The covariance of spin dynamics with respect to the rotation of
quantization axis strongly resembles the gauge covariance of QCD.
Thus, it was argued long ago \cite{Mro87} that the QCD collision terms
are of the Waldmann-Snider type \cite{Wal57} known from the studies
of spin systems. More recently, guided by the same analogy, the
Waldmann-Snider transport equations have been used to compute color 
conductivity of the quark-gluon plasma \cite{Arnold:1998cy}, as well 
as other transport coefficients \cite{Arnold:2000dr,Arnold:2003zc}.

Once the collision terms of transport equations are known, the problem of
finding the state of local equilibrium is well posed, see {\it e.g.} 
\cite{Gro80}. Namely, one looks for a configuration which maximizes the 
local entropy. In fact, such a configuration can be also found without 
a detailed knowledge on the collision terms. One only needs the so-called 
collisional invariants - the conditions  obeyed by the collision 
terms, coming from the conservations laws. In such an approach, already 
followed in \cite{Dyrek:1986vv,Mrowczynski:1989bv}, we, however, gain no 
information about the time scale corresponding to the local equilibrium sate. 
We also do not know whether the local equilibrium configuration dictated by
the collisional invariants is the most general maximum entropy state.
To answer these questions an explicit form of the collision terms is
required. Then, one looks for a configuration that cancels the collision
terms. 

In this paper we follow both approaches. After introducing the kinetic 
theory of the quark-gluon plasma in Sec. \ref{kinetic-theory}, we find 
in Sec. \ref{loc-cons} the local equilibrium state provided by the 
collisional invariants. Then, we select the most probable binary
interactions and we derive in Sec. \ref{loc-col} the local equilibrium 
functions which cancel the Waldmann-Snider collision terms corresponding
to these dominant processes. The derivation requires solving a whole
set of rather complicated matrix equations. To simplify the analysis, 
we consider particles obeying classical statistics, although we believe
that the physical picture emerging from our analysis is not much changed
when  quantum statistics is incorporated. The local equilibrium states,
which come from the approaches of Secs.\ref{loc-cons} and \ref{loc-col},
  are colorful and their color structure 
is exactly the same. However, the baryon chemical potential of (anti-)quarks 
and the scalar chemical potential of gluons remain unconstrained by the 
dominant processes. The constraints provided by the collisional invariants 
only appear when the subleading quark-antiquark annihilation/creation and 
gluon number non-conserving processes are included. To better understand
properties of the colorful local equilibrium, we derive in 
Sec. \ref{color-hydro} the resulting hydrodynamic equations. Finally, 
we consider the applicability of our results and briefly discuss possible 
processes responsible for the color neutralization in the quark-gluon 
plasma. Some formulas of the SU($N_c$) generators are collected in 
the Appendix.

Throughout the paper (except Eqs.~(\ref{3-vectors}-\ref{Euler-non-rel}) 
where $c$ is restored) we use the natural units with $c=\hbar=k_B =1$ and 
the metric $(1,-1,-1,-1)$.


\section{Kinetic theory of the quark-gluon plasma}
\label{kinetic-theory}


In this section we discuss the transport theory of quarks and
gluons \cite{Elze:1989un,Mrowczynski:np}. The SU($N_c$) gauge group 
is left unspecified but we pay a particular attention to the cases 
$N_c =2 $ and $N_c =3$, for their possible applications to the different 
high temperature phases of the Standard Model. Generically speaking, we 
call gluons the particles associated to the vector bosons of SU($N_c$), 
which carry charge in the adjoint representation, and we call quarks 
or antiquarks the particles with the charge in the fundamental 
representation. We will also call parton any of those particles.


\subsection{Distribution functions and transport equations}
\label{sec-transport}


The distribution function of quarks $Q(p,x)$ is a hermitian
$N_c\times N_c$ matrix in color space (for a SU($N_c$) color
group); $x$ denotes the space-time quark coordinate and $ p$
its momentum, which is not constrained by the mass-shell condition.
The spin of quarks and gluons is taken into account as an internal
degree of freedom. The distribution function transforms under
a local gauge transformation $U$ as
\begin{equation}
\label{Q-transform}
Q( p,x) \rightarrow U(x) \, Q( p,x) \, U^{\dag }(x) \; ,
\end{equation}
that is, it transforms covariantly in the fundamental representation.
Here and in the most cases below, the color indices are suppressed.
The distribution function of antiquarks, which we denote by $\bar Q(p,x)$,
is also a hermitian $N_c\times N_c$ matrix in color space, which in a
natural way should transform covariantly in the conjugate fundamental
representation. However, we will express the antiquark distribution
function in the same representation as quarks throughout, and then it
transforms according to Eq.~(\ref{Q-transform}). The distribution function 
of (hard) gluons is a hermitian $(N_c^2-1)\times (N_c^2-1)$ matrix, which
transforms as
\begin{equation}
\label{G-transform}
G( p,x) \rightarrow {\cal U}(x) \: G( p,x) \:{\cal U}^{\dag }(x) \;,
\end{equation}
where
\be
{\cal U}_{ab}(x) = 2{\rm Tr}\bigr[\tau^a U(x) \tau^b U^{\dag }(x)] \ ,
\ee
with $\tau^a ,\; a = 1,...,N_c^2-1$ being the SU($N_c$) group generators
in the fundamental representation with ${\rm Tr} (\tau^a \tau^b) = \frac12
\delta^{ab}$. We note that ${\cal U}^T = {\cal U}^{-1} = {\cal U}^{\dag}$.
Therefore, not only $G$ but also $G^T$ transforms covariantly {\it i.e.}
\be
G^T(p,x) \rightarrow {\cal U}(x) \: G^T(p,x) \: {\cal U}^{\dag }(x) \;.
\ee

The color current is expressed in the fundamental representation as
\be
\label{col-current}
j^{\mu }(x) = -\frac{g}{2} \int dP \; p^\mu \;
\Big[ Q( p,x) - \bar Q ( p,x) 
- {1 \over N_c}{\rm Tr}\big[Q( p,x) - \bar Q ( p,x)\big] 
+  2 \tau^a {\rm Tr}\big[T^a G(p,x) \big]\Big] \; ,
\ee
where the momentum measure
\be
dP \equiv \frac{d^4p}{(2 \pi)^3} 2\Theta(p_0)\, \delta(p^2)
\ee
takes into account the mass-shell condition  $p_0 = |{\bf p}|$.
Throughout the paper, we neglect the quark masses, although those
might be easily taken into account by modifying the mass-shell
constraint in the momentum measure.  A sum over helicities, two 
per particle, and over quark flavors $N_f$ is understood in 
\eq\nr{col-current}, even though it is not explicitly written 
down. The SU($N_c$) generators in the adjoint representation are 
expressed through the structure constants  $T^a_{bc} = -i f_{abc}$, 
and are normalized as ${\rm Tr}[T^aT^b]= N_c \delta^{ab}$. 
The current can be decomposed as $j^\mu (x) = j^\mu_a (x) \tau^a$ 
with $j^\mu_a (x) = 2 {\rm Tr} (\tau_a j^\mu (x))$. 

Gauge invariant quantities are given by the traces
of the distribution functions. Thus, the baryon current and the
energy-momentum tensor read
\ba
\label{bar-current}
b^{\mu }(x) &=& {1 \over 3} \int dP \; p^{\mu} \;
{\rm Tr}\Big[ Q(p,x) - \bar Q (p,x) \Big] \; , \\[2mm]
\label{en-mom}
t^{\mu \nu}(x) &=&  \int dP \; p^{\mu} p^{\nu} \;
{\rm Tr}\Big[ Q(p,x) + \bar Q (p,x) + G(p,x) \Big] \; ,
\ea
where we use the same symbol ${\rm Tr}[\cdots]$ for the trace
in the fundamental and adjoint representations.

The entropy flow is defined as \cite{Dyrek:1986vv}
\ba \label{entropy}
s^{\mu }(x) =  - \int dP \; p^{\mu} \; {\rm Tr}\Big[
Q {\rm ln}Q  + (1-Q) {\rm ln}(1-Q)
&+& \bar Q {\rm ln}\bar Q
+ (1-\bar Q) {\rm ln}(1- \bar Q) \\ [2mm] \nonumber
&+& G {\rm ln}G - (1 + G) {\rm ln}(1+ G) \nonumber \Big] \;.
\ea
If the effects of quantum statistics are neglected, \eq\nr{entropy}
simplifies to
\be \label{entropy-class}
s^{\mu }(x) =  - \int dP \; p^{\mu} \; {\rm Tr}\Big[
Q ({\rm ln}Q -1) + \bar Q ({\rm ln}\bar Q -1)
+ G ({\rm ln}G -1) \Big] \;.
\ee

The distribution functions of quarks and gluons satisfy the transport
equations:
\begin{mathletters}
\label{transport}
\ba
p^{\mu} D_{\mu}Q(p,x) + {g \over 2}\: p^{\mu}
\left\{ F_{\mu \nu}(x), \partial^\nu_p Q(p,x) \right\}
&=& C[Q,\bar Q,G] \;,
\label{transport-q}  \\ [2mm]
p^{\mu} D_{\mu}\bar Q(p,x) - {g \over 2} \: p^{\mu}
\left\{ F_{\mu \nu}(x), \partial^\nu_p \bar Q(p,x)\right\}
&=& \bar C[Q,\bar Q,G]\;,
\label{transport-barq} \\ [2mm]
p^{\mu} {\cal D}_{\mu}G(p,x) + {g \over 2} \: p^{\mu}
\left\{ {\cal F}_{\mu \nu}(x), \partial^\nu_p G(p,x) \right\}
&=& C_g[Q,\bar Q,G]\;,
\label{transport-gluon}
\ea
\end{mathletters}
where  $g$ is the QCD coupling constant, $\{...,...\}$ denotes the
anticommutator and $\partial^\nu_p$ the four-momentum derivative; the
covariant derivatives $D_{\mu}$ and ${\cal D}_{\mu}$ act as
$$
D_{\mu} = \partial_{\mu} - ig[A_{\mu}(x),...\; ]\;,\;\;\;\;\;\;\;
{\cal D}_{\mu} = \partial_{\mu} - ig[{\cal A}_{\mu}(x),...\;]\;,
$$
with $A_{\mu }$ and ${\cal A}_{\mu }$ being four-potentials
in the fundamental and adjoint representations, respectively:
$$
A^{\mu }(x) = A^{\mu }_a (x) \tau^a \;,\;\;\;\;\;
{\cal A}^{\mu }(x) = T^a A^{\mu }_a (x) \; .
$$
The stress tensor in the fundamental representation is 
$F_{\mu\nu}=\partial_{\mu}A_{\nu} - \partial_{\nu}A_{\mu}
-ig [A_{\mu},A_{\nu}]$, while  ${\cal F}_{\mu \nu}$ denotes the field
strength tensor in the adjoint representation. The collision terms
$C, \bar C$ and $C_g$ are discussed in detail in the next subsections.

Let us finally mention that in the transport theory framework one can 
consider two different physical situations: 1) the gauge fields entering 
into the transport equations (\ref{transport}) are external, not due to 
the plasma constituents; 2) the gauge fields can be generated self-consistently
by the quarks and gluons. In the latter case, one also has to solve the
Yang-Mills equation
\be
\label{yang-mills}
D_{\mu} F^{\mu \nu}(x) = j^{\nu}(x)\; ,
\ee
where the color current is given by \eq\nr{col-current}.

\subsection{Decomposition of the distribution functions and associated
transport equations}
\label{App.decom}

The parton distribution function $N$ is essentially the statistical
average of the Wigner transform of the product of two field operators
representing quarks or gluons \cite{Elze:1989un}. If the parton carries 
color charge in a representation $R$, then the distribution function $N$ 
transforms under gauge transformations as $\bar R \otimes R$, where 
$\bar R$ is the representation conjugate to $R$.

In the SU(2) group, the products of the fundamental (${\bf 2}$) and
adjoint (${\bf 3}$) representations decompose into irreducible
representations as
\ba
\label{q-decomSU(2)}
{\bf 2} \otimes {\bf 2} & = &  {\bf 1} \oplus {\bf 3} \;, \\
{\bf 3} \otimes {\bf  3} & = &  {\bf 1} \oplus {\bf 3} \oplus  {\bf 5} \;.
\label{g-decomSU(2)}
\ea
As known, the conjugate and direct fundamental representations of SU(2)
are equivalent to each other. The decomposition of the products of the
fundamental (${\bf 3}$) and adjoint (${\bf 8}$) representations of the
SU(3) group are
\ba
\label{q-decomSU(3)}
{\bf 3} \otimes {\bf \bar 3} & = &  {\bf 1} \oplus {\bf 8} \;, \\
{\bf 8} \otimes {\bf  8} & = &  {\bf 1} \oplus {\bf 8} \oplus  {\bf 8}
\oplus {\bf 10} \oplus {\bf {\overline{10}}} \oplus {\bf 27}   \;.
\label{g-decomSU(3)}
\ea

The above decompositions show that the distribution functions of quarks
and antiquarks are uniquely specified by their singlet and adjoint 
components.  Thus, the functions can be written as
\begin{mathletters}
\label{quprojec-component}
\ba 
Q(p,x)  & = & \frac{1}{N_c}  q_0 (p,x) + q^a(p,x) \tau^a \; , \\
\bar Q(p,x)  &  = & \frac{1}{N_c}  \bar q_0 (p,x) + \bar q^a(p,x) \tau^a \;,
\ea
\end{mathletters}
where
\begin{mathletters}
\label{projec-component}
\ba
q_0 (p,x) & =  &{\rm Tr}[Q(p,x)] \ , \qquad
 q_a(p,x) = 2 {\rm Tr}[ \tau^a Q(p,x)] \ , \\
\bar q_0 (p,x) & =  &{\rm Tr}[\bar Q(p,x)] \ , \qquad
 \bar q_a(p,x) = 2 {\rm Tr}[ \tau^a \bar Q(p,x)] \;. 
\ea
\end{mathletters}

From Eq.~(\ref{transport-q}) it is possible to deduce  a set of coupled
equations for the colored and colorless components of the quark
distribution function which read
\begin{mathletters}
\label{projec-transport}
\ba
p^\mu \partial_\mu  q_0(p,x)
&+& \frac{g}{2}\, p^\mu F_{\mu \nu}^a(x)
\frac{\partial q_a(p,x)}{\partial p_\nu}  =  {\rm Tr}[C] \; , \\ [2mm]
p^\mu  D_\mu ^{ab} q_b (p,x)
&+& \frac{g}{2}\, d_{abc} p^\mu F_{\mu \nu}^b(x)
\frac{\partial q_c(p,x)}{\partial p_\nu}
+ \frac{g}{N_c}\, p^\mu F_{\mu \nu}^a(x)
\frac{\partial q_0 (p,x)}{\partial p_\nu} =  2 {\rm Tr}[\tau^a C] \;,
\ea
\end{mathletters}
where $d_{abc}$ are the totally symmetric structure constants of
SU($N_c$) and $D_\mu^{ac} = \partial_\mu \delta^{ac} + g f_{abc} A_\mu ^b$. 
The projected equations, which can be also written for antiquarks,
show that transport phenomena of colorless and colored components
are coupled beyond the lowest order in the gauge coupling constant. 

From the decompositions (\ref{g-decomSU(2)},\ref{g-decomSU(3)}) it 
is clear that the singlet and adjoint components are not enough to 
fully describe the gluon distribution function. For gluons one also 
needs components in higher dimensional representations. Below, we 
present a way to uniquely characterize the gluon distribution 
function in terms of its fully symmetric and antisymmetric components
for the SU(2) gauge theory.

We first express $G(p,x)$ as
\be
\label{gluonbasis}
G(p,x) = {\cal G}_{ab} (p,x) T^a T^b \; ,
\ee
which uses as a basis for $3 \times 3$ hermitian matrices the set of 
9 independent matrices $ T^a T^b$. We note that both $G$ and ${\cal G}$
are $3 \times 3$ matrices which are related to each other as
\be
G_{ab}(p,x) = \delta^{ab} {\cal G}_{cc} (p,x) -{\cal G}_{ba}(p,x) \;.
\ee

Expressing the product of $T^aT^b$ as
\be
T^a T^b = \frac 12 [T^a, T^b] + \frac 12 \{T^a,T^b \} \;,
\ee
and taking into account that the commutator is proportional to $T^c$,
instead of Eq.~(\ref{gluonbasis}) we write 
\be \label{gluonbasis2}
G(p,x) = \frac 12 g_a (p,x) \: T^a 
+ g_{ab} (p,x) \: \frac 12 \{T^a, T^b\} \;,
\ee
where
\be \label{G-vs-g}
g_a (p,x) = i f^{abc} \: {\cal G}_{cb} (p,x) \;, \qquad
g_{ab} (p,x) = \frac 12 \left({\cal G}_{ab} (p,x) 
+ {\cal G}_{ba} (p,x) \right) \;.
\ee
The equation (\ref{gluonbasis2}) can be also written as
\be
G_{ab}(p,x) = -if^{abc} g_c(p,x) + \delta^{ab} g_{cc} (p,x) 
- g_{ab} (p,x) \;.
\ee 
Thus, according to the decomposition in Eq.~(\ref{g-decomSU(2)}),
the antisymmetric components of $G$ correspond to the representation 
${\bf 3}$, while the 6 symmetric components correspond to the ${\bf 5}$ 
and ${\bf 1}$, the last one being the trace. Because of the Casimir 
constraint, $T^a T^a = 2$, the singlet component can be obtained from 
the symmetric part $g_{ab}$ {\it i.e.}
\be
g_0(p,x)  \equiv {\rm Tr} [G(p,x)] = 2\: g_{aa} (p,x) \;.
\ee

The transport equations obeyed by $g_0$, $g_a$ and $g_{ab}$ 
are found multiplying Eq.~(\ref{transport-gluon}) by the unity, 
$T^a$ and $\{T^a, T^b\}/2$, respectively, and taking the trace.
Using the relations (\ref{tracesSU2}), we get
\begin{mathletters}
\label{projec-transport-gSU2}
\ba
\label{projec-transport-g-singSU2}
p^\mu \partial_\mu  g_0(p,x)
&+& g\, p^\mu F_{\mu \nu}^a(x)
\frac{\partial g_a(p,x)}{\partial p_\nu}  =  {\rm Tr}[C_g] \;, 
\\
\label{projec-transport-g-antiSU2}
 p^\mu  D_\mu ^{ab} g_b (p,x)
&+& g  p^\mu  F_{\mu \nu}^b(x)
\left( \frac 12 \: \delta^{ab}
\frac{\partial g_0 (p,x)}{\partial p_\nu} + 
\frac{\partial g_{ab} (p,x)}{\partial p_\nu} \right)
 = {\rm Tr}[T^a C_g] \;, \\
\label{projec-transport-g-symSU2}
p^\mu ( D_\mu)^{ac}_{bd}\: g_{cd} (p,x)
&+& \frac{g}{4}  \, p^\mu \left( F_{\mu \nu}^a(x)
\frac{\partial g_{b} (p,x)}{\partial p_\nu} + F_{\mu \nu}^b(x)
\frac{\partial g_{a} (p,x)}{\partial p_\nu} \right)
 = \frac 12 {\rm Tr}[\{T^a,T^b\} C_g] 
- \frac 12 \delta^{ab} {\rm Tr}[C_g] \;,
\ea
\end{mathletters}
where 
\be
(D_\mu)^{ac}_{bd} =\partial_\mu \delta^{ac} \delta^{bd}
+ g f^{aec} \delta^{bd} \: A_\mu^e 
+ g f^{bed} \delta^{ac} \: A_\mu^e
\ee
is the covariant derivative acting on a tensor of rank 2. Note that
multiplying the last equation by $\delta^{ab}$, we get, as expected, 
the equation for $g_0$.

For SU(3), or SU($N_c$) in general, the decomposition of the gluon
distribution function into irreducible representations 
and the equations obeyed by every component have 
a much more involved structure, and they will not be discussed here.


\subsection{Waldmann-Snider Collision terms}
\label{wald-sec}


The transport equations for the quark-gluon plasma (\ref{transport})
have been written down without specifying the collision terms.
Unfortunately, a complete derivation of $C$, $\bar C$ and $\bar C_g$
is still lacking, as already mentioned in the Introduction. However, using the
analogy with the spin systems one can justify the use of the 
Waldmann-Snider collision terms. The main characteristic of these
collision terms is that they depend on the scattering amplitudes 
rather than on the collisional cross sections, as it happens in the
usual Boltzmann equation.

Let us discuss the general structure of a collision term for a 
system of particles carrying quantum color charges. The most probable processes
are binary collisions $(p,r;p_1,s) \leftrightarrow (p',t;p_1',u)$ where
$p,p_1,p',p_1'$ denote the momenta and $r,s,t,u$ colors, in the fundamental
or adjoint representation, of interacting partons. We denote by $N(p,x)$
the generic distribution function of the partons - quarks or gluons. The
Waldmann-Snider collision term, which enters the kinetic equation of $N$,
is of the form \cite{Gro80}:
\be \label{Wald}
C[N,N_1,N',N_1']  =  \int dP' dP_1'dP_1 \;
(2 \pi)^4 \delta^{(4)} (p+p_1 -p'-p_1')
 \Big[ \frac{1}{2} \left\{1 \pm N, {\cal I}_+ \right\}
- \frac{1}{2} \left\{N, {\cal I}_- \right\} \Big] \;,
\ee
where we have used a rather common notation $N \equiv N(p,x)$,
$N_1 \equiv N(p_1,x)$, $N' \equiv N(p',x)$, and $N_1' \equiv N(p_1',x)$.
The first term, which represents a gain term, is given by
\be \label{gainWS}
{\cal I}_+^{r \bar r} =
M_{rstu}(p,p_1;p',p'_1) \;
M^*_{\bar r \bar s \bar t \bar u}(p,p_1;p',p'_1) \;
N^{t \bar t}(p',x) \; N^{u \bar u} (p'_1,x) \;
\left( 1 \pm N(p_1,x) \right)^{\bar s s} \;,
\ee
while the second one is a loss term defined as
\be \label{lossWS}
{\cal I}_-^{r \bar r} =
M_{rstu}(p,p_1;p',p'_1)\;
M^*_{\bar r \bar s \bar t \bar u}(p,p_1;p',p'_1) \;
N^{\bar s  s}(p_1,x) \; \left( 1 \pm N(p',x) \right)^{t \bar t } \;
\left( 1 \pm N(p'_1,x) \right)^{u \bar u} \;.
\ee
$M_{rstu}$ represents the scattering amplitude associated with the
collision process under consideration. The double sign $\pm$
reflects the fermionic character of quarks and bosonic of gluons.
We have used here the compact notation of ref. \cite{Arnold:1998cy}.

For the consistency of the theory, it is necessary to prove that
the Waldmann-Snider collision terms transform covariantly under
a gauge transformation, in the same way as the left hand sides of
the transport equations (\ref{transport}) do. It is difficult to 
check this gauge covariance in full generality without specifying 
the scattering process  and the corresponding  scattering amplitudes.
For all the cases we are going to consider, the gauge covariance of the
Waldmann-Snider collision term holds as the distribution functions transform
covariantly (see Eqs.~(\ref{Q-transform}) and (\ref{G-transform})),
and the scattering amplitudes, stripped of the color generators,
are gauge invariant.  We will briefly come back to this point in 
Sec. \ref{loc-col}.


\subsection{Conservation laws and entropy production}
\label{cons-sec}


As well known, the collision terms should satisfy certain relations due
to the conservation laws. In our case, the laws are: the baryon charge
conservation
\be \label{bar-cons}
\partial_\mu b^\mu(x) = 0 \;,
\ee
the energy-momentum conservation
\be \label{en-mom-cons}
\partial_\mu t^{\mu \nu}(x)
+ 2{\rm Tr}[j_\sigma(x)\; F^{\sigma \nu}(x) ] = 0 \;,
\ee
and the covariant conservation of the color current
\be \label{col-cons}
D_\mu j^\mu(x) = 0 \;.
\ee

Let us derive the relations constraining the collision terms which follow
from Eqs.~(\ref{bar-cons},\ref{en-mom-cons},\ref{col-cons}).
Using the transport equation, one finds from the definition \nr{bar-current}
$$
\partial_\mu b^\mu(x) = {1 \over 3} \int dP \; {\rm Tr}[ C - \bar C]
-{g \over 3} \int dP p^\sigma \; {\rm Tr}[F_{\sigma \nu} \partial^\nu_p
(Q + \bar Q)] \;.
$$
Now, one performs partial integration of the second term in the r.h.s.
Assuming that the distribution functions vanish at infinite momentum
and observing that $g^{\mu \nu} F_{\mu \nu} = 0$, one finds that the
term equals zero. Therefore, the baryon current conservation
\nr{bar-cons} provides
\be 
\label{bar-cons-inv}
\int dP \; {\rm Tr}[ C - \bar C] = 0 \;.
\ee

In analogous way, one finds that the energy-momentum conservation
\nr{en-mom-cons} implies
\be 
\label{en-mom-cons-inv}
\int dP \;p^\mu \; {\rm Tr}[ C + \bar C + C_g] = 0 \;,
\ee
while the covariant conservation of the color current leads to
\be \label{col-cons-inv}
\int dP \; \big[ C - \bar C + 2 \tau^a {\rm Tr}[ T^a C_g] \big]= 0 \;,
\ee
where we have taken into account the relation \nr{bar-cons-inv}.

Let us now discuss the entropy production. We neglect here the effects
of quantum statistics, and consequently start with the definition
\nr{entropy-class}. Following the derivation of
Eqs.~(\ref{bar-cons-inv},\ref{en-mom-cons-inv},\ref{col-cons-inv}),
one finds
\ba \label{entro-prod0}
\partial_\mu s^\mu(x) &=& - \int dP \;
{\rm Tr}[ C {\rm ln}Q + \bar C {\rm ln}\bar Q+ C_g {\rm ln}G]
\\ [2mm] \nonumber
&-& {g \over 2} \int dP \;p^\mu \;
{\rm Tr}\big[ \{F_{\mu \nu},Q \} \partial^\nu_p {\rm ln}Q
-\{F_{\mu \nu},\bar Q \} \partial^\nu_p {\rm ln}\bar Q
+ \{{\cal F}_{\mu \nu},G \} \partial^\nu_p {\rm ln}G \big] \;,
\ea
where the partial integration has been once performed and it has
been observed that
$$
{\rm Tr} \Big[ [A^\mu,Q] \; {\rm ln}Q \Big] = 0 \;,
$$
and that the analogous equalities hold for $\bar Q$ and $G$. Assuming 
that $Q$ and $\partial^\nu_p Q$ commute with each other {\it i.e.}
\be 
\label{diagonalcondition}
[Q ,\partial^\nu_p Q ] = 0 \;,
\ee
one shows that $\partial^\nu_p {\rm ln}Q  = Q^{-1} \partial^\nu_p Q $.
Using the condition \nr{diagonalcondition} and the similar ones for
$\bar Q$ and $G$, one proves that the second term in r.h.s. of
\eq\nr{entro-prod0} vanishes after one more partial integration.
Then, we get 
\be \label{entro-prod}
\partial_\mu s^\mu(x) = - \int dP \;
{\rm Tr}[ C {\rm ln}Q + \bar C {\rm ln}\bar Q+ C_g {\rm ln}G] \;.
\ee
According to Eq.~(\ref{entro-prod}), the entropy of the quark-gluon 
system is produced due to the collisions. If the commutation condition (\ref{diagonalcondition}) is relaxed, the second term in r.h.s. of 
\eq\nr{entro-prod0} does not vanish, and we arrive to a paradoxical 
result that the mean-field dynamics does not conserve the entropy.

A local equilibrium configuration is achieved when there is no entropy
production, {\it i.e.} $\partial_\mu s^\mu(x) = 0$. This equation is 
of very complicated structure and it has two classes of solutions. 
The first one cancels the collision terms but to get it the collision 
terms have to be specified. The second class appears due to the 
conservation laws, {\it i.e.}, because of the relations 
(\ref{bar-cons-inv},\ref{en-mom-cons-inv}) and (\ref{col-cons-inv}). 
In the remaining part of this article, we will study the two sets of 
solutions.


\section{Local equilibrium from the conservation laws}
\label{loc-cons}


In this section we discuss, following \cite{Dyrek:1986vv,Mrowczynski:1989bv},
consequences of the conservation laws
(\ref{bar-cons},\ref{en-mom-cons},\ref{col-cons}). Specifically,
we obtain the local equilibrium configuration which is found as
a solution of the equation
\be \label{loc-eq-eq}
\int dP \; {\rm Tr}[ C {\rm ln}Q + \bar C {\rm ln}\bar Q
+ C_g {\rm ln}G]  = 0 \;,
\ee
due to the relations
(\ref{bar-cons-inv},\ref{en-mom-cons-inv},\ref{col-cons-inv}).

One easily constructs the local equilibrium distribution function
out of the collision invariants. Indeed, one shows using
Eqs.~(\ref{bar-cons},\ref{en-mom-cons},\ref{col-cons}) that
\eq\nr{loc-eq-eq} is solved if
\begin{mathletters}
\label{loc-eq}
\ba \label{loc-eq-Q}
Q_{\rm eq}(p,x) &=& {\rm exp}\Big[- \beta (x)
\big(u_\nu (x) p^\nu - \mu_b(x) - \widetilde \mu (x) \big)\Big] \;,
\\ [2mm] \label{loc-eq-barQ}
\bar Q_{\rm eq}(p,x) &=& {\rm exp}\Big[-\beta (x)
\big(u_\nu (x) p^\nu + \mu_b(x) + \widetilde \mu (x) \big)\Big] \;,
\\ [2mm] \label{loc-eq-G}
G_{\rm eq}(p,x) &=& {\rm exp}\Big[-\beta (x)
\big(u_\nu (x) p^\nu - \widetilde \mu_g (x) \big)\Big] \;,
\ea
\end{mathletters}
where $\beta (x)$, $u^{\nu}(x)$ and $\mu_b(x)$ are, respectively, the
inverse temperature, hydrodynamic velocity and baryon chemical potential 
which are all scalars in color space. The color chemical potentials
$\widetilde \mu $ and $\widetilde \mu _g$ are  hermitian matrices
$N_c \times N_c$ for quarks and $(N_c^2-1)\times (N_c^2-1)$ for gluons.
They are gauge dependent variables, which transform as
\be
\widetilde \mu(x) \rightarrow U(x) \, \widetilde \mu(x) \, U^{\dag }(x) 
\ , \qquad
\widetilde \mu _g (x) \rightarrow 
{\cal U}(x) \: \widetilde \mu _g(x) \: {\cal U}^{\dag }(x) \;.
\ee
In general, $\widetilde \mu$ can be expressed as 
$\widetilde \mu = \mu_0 + \mu_a \tau^a$. However, the singlet component 
$\mu_0$ is already singled out as a baryon chemical potential $\mu_b$. 
Therefore, we write down $\widetilde \mu = \mu_a \tau^a$.
Consequently, the color chemical potential $\widetilde \mu$
is not only hermitian but also traceless. The covariant conservation
of the color current provides the relation
\be
\label{relat-qgmu}
\widetilde \mu _g = 2 T^a {\rm Tr}[\tau^a \widetilde \mu ] 
= \mu_a  T^a \;,
\ee
which implies that $\widetilde \mu _g$ is also traceless. 
The baryon and color chemical potentials occur in \eqs(\ref{loc-eq})
because of the conservation laws of baryon number and color charge, 
respectively. The temperature and hydrodynamic velocity are
related to the energy-momentum conservation.
  
The local equilibrium state described by \eqs(\ref{loc-eq}) is {\em not}
color neutral. Substituting the distribution functions (\ref{loc-eq}) 
into \eq\nr{col-current} one finds the color current as
\be \label{loc-cur}
j^{\mu} = - g \,{T^3 \over \pi^2} \; u^{\mu}
\Big[ N_f \Big( e^{\beta \mu_b} \big( e^{\beta \tilde \mu} -
{1 \over N_c}\, {\rm Tr}[e^{\beta \tilde \mu}] \big)
- e^{-\beta \mu_b} \big( e^{-\beta \tilde \mu} -
{1 \over N_c}\, {\rm Tr}[e^{-\beta \tilde \mu}]\big) \Big)
+ 2 \tau^a {\rm Tr}[T^a e^{\beta \tilde \mu_g}] \Big] \;,
\ee
where $T$, $u^\mu$, $\mu_b$, $\widetilde\mu$, and $\widetilde\mu_g$
are functions of $x$. The fact that the color current is finite does
not imply that the system as a whole carries a finite color charge. 
We note that the $x-$dependence of the color chemical potentials,
which enter the solutions (\ref{loc-eq}), is not specified.
Therefore, it can be always chosen in such a way that the total
color charge defined as $\int d^3x\,j^0$ vanishes.

The derivation of the local distribution function based on the 
collisional invariants tells nothing about the time scales 
when the colorful configuration (\ref{loc-eq}) exists. To get 
such an information the collision terms have to be specified.
This is discussed in the next sections.

The equilibrium solutions (\ref{loc-eq}) are given in an arbitrary gauge.
It is often useful to work in a gauge where the quark and antiquark chemical 
potentials are diagonal. Then,
\be
\widetilde \mu = \mu^d \tau^d \ , \qquad \widetilde \mu _g = \mu^d T^d \ ,
\ee
where $\tau^d$ and $T^d$ are the fundamental and adjoint generators
of the Cartan subalgebra of SU($N_c$) ($d=3$ for SU(2) and $d=3,8$ for
SU(3)). In this gauge one has, as will be seen below, well-defined numbers
of quarks and antiquarks of a certain color. And then, the physical
meaning of the color chemical potentials becomes transparent.


\subsection{Diagonal gauge for SU(2)}


Using the explicit form of $\tau^3 = \sigma^3/2$, where $\sigma^3$
is the Pauli matrix, the singlet and the (non-vanishing) adjoint 
components (see Eqs.~(\ref{quprojec-component},\ref{projec-component})) 
of the quark and antiquark distribution functions of local equilibrium 
(\ref{loc-eq}) are found as
\begin{mathletters} 
\label{loc-eq-q-diag-SU2} 
\ba
q_0 (p,x) = q_{\uparrow}(p,x) + q_{\downarrow}(p,x)  \;,
 \qquad   q_3 (p,x) = q_{\uparrow}(p,x) - q_{\downarrow}(p,x) \;,
 \\ [2mm]  
\bar q_0(p,x) =  \bar q_{\uparrow}(p,x) + \bar q_{\downarrow} (p,x)   \;,
 \qquad   \bar q_3 (p,x) = 
\bar q_{\uparrow}(p,x) - \bar q_{\downarrow}(p,x) \;,
\ea
where the scalar functions $q_{\uparrow \downarrow}$ and 
$\bar q_{\uparrow \downarrow}$ are 
\ba
\label{qlocalup}
q_{\uparrow \downarrow} (p,x) &\equiv& 
\exp{\left[- \beta(x)\left( u^\mu(x) p_\mu -\mu_b (x) 
                   \mp \frac 12 \mu_3 (x) \right) \right] }  \;, 
\\ [2mm] \label{barlocalup}
\bar q_{\uparrow \downarrow} (p,x) &\equiv& 
\exp{\left[- \beta(x)\left( u^\mu(x) p_\mu
           +\mu_b (x) \pm \frac 12 \mu_3 (x) \right) \right] }  \;. 
\ea
\end{mathletters}
While the generator $\tau^3$ is diagonal, $T^3$ is not. To derive
the expressions for gluons one has to observe that $(T^3)^2$ is the diagonal
matrix with $1,1,0$ on the diagonal. Consequently, $(T^3)^n = T^3$
when $n = 1,3,5 \dots$ and $(T^3)^n = (T^3)^2$ when $n = 2,4,6 \dots$.
Thus, the non-vanishing components (\ref{gluonbasis2}) of the gluon 
distribution function (\ref{loc-eq-G}) are
\begin{mathletters}
\label{loc-eq-g-diag-SU2} 
\ba
g_0 (p,x) &=& g_{\Uparrow}(p,x) + g_{\Rightarrow} (p,x) + g_{\Downarrow}(p,x) 
\;, \\ [2mm]
g_3(p,x) &=& g_{\Uparrow}(p,x) - g_{\Downarrow} (p,x) \;, 
\\ [2mm] 
g_{11}(p,x) &=& g_{22}(p,x) = {1 \over 2}\, g_{\Rightarrow}(p,x) \;,
\\ [2mm]
g_{33}(p,x) &=& {1 \over 2}\, \Big(g_{\Uparrow}(p,x) 
+ g_{\Downarrow}(p,x) - g_{\Rightarrow} (p,x) \Big) \;,
\ea
where the functions $g_{\Uparrow \Downarrow}$ and $g_{\Rightarrow}$ are
\be 
\label{glocalup}
g_{\Uparrow \Downarrow} (p,x) \equiv 
\exp{\left[- \beta(x)\left( u^\mu(x) p_\mu 
\mp  \mu_3 (x)  \right) \right] }  \;, \;\;\;\;\;\;\;\;\;
g_{\Rightarrow} (p,x) \equiv \exp{\left[- \beta(x) u^\mu(x) p_\mu \right] }  
\;.
\ee
\end{mathletters}
In the diagonal gauge, a finite value of the color chemical potential simply 
means that the populations of quarks, antiquarks and gluons of different 
colors are not the same.

Using the distribution functions in the form 
(\ref{loc-eq-q-diag-SU2},\ref{loc-eq-g-diag-SU2}), the color current 
(\ref{loc-cur}) can be written as
\be
j^{\mu} = - 4g \,{T^3 \over \pi^2} \; u^{\mu}
\Big[N_f {\rm ch}(\beta \mu_b) \; {\rm sh}(\beta \mu_3/2 )
+ {\rm sh}(\beta \mu_3 )\Big] \tau^3 \;.
\ee


\subsection{Diagonal gauge for SU(3)}


The local equilibrium solutions for the SU(3) plasma can be also written 
in the diagonal gauge. However, the formulas are not that simple as for 
the SU(2) case. We take the generators in the fundamental representation
as $\tau^a = \lambda^a/2$, where $\lambda^a$ are the Gell-Mann matrices. 
The matrices $\lambda^3$ and $\lambda^8$ are diagonal with the elements 
1, $-1$, 0 and $1/\sqrt{3}$, $1/\sqrt{3}$, $-2/\sqrt{3}$, respectively,
along the diagonal. With a color chemical potential in the directions 
$a=3$ and $a=8$ one can then easily evaluate the singlet and (non-vanishing) 
adjoint components of $Q_{\rm eq}$ and $\bar Q_{\rm eq}$, which we write 
in terms of the distributions functions of red, blue and green quarks 
and antiquarks.  Here, we have taken the convention to assign the first, 
second and third rows/columns of the Gell-Mann matrices to the red, blue 
and green colors, respectively. A simple evaluation leads to
\begin{mathletters} 
\label{loc-eq-diag-SU3} 
\ba
q_0 (p,x) &=& q_{\rm red}(p,x) + q_{\rm blue}(p,x) + q_{\rm green}(p,x) 
  \;, \qquad   
q_3 (p,x) = q_{\rm red}(p,x) - q_{\rm blue} (p,x) \;,
 \\ [2mm]  
q_8(p,x) &=&  \frac{1}{\sqrt{3}} \Big( q_{\rm red} (p,x) +  q_{\rm blue}(p,x) 
- 2 q_{\rm green} (p,x)  \Big)  
\;, \\ [4mm]
\bar q_0 (p,x) &=& \bar q_{\rm red}(p,x) +\bar  q_{\rm blue}(p,x) 
                  +\bar q_{\rm green}(p,x) 
  \;, \qquad  
\bar q_3 (p,x) = \bar q_{\rm red}(p,x) -\bar  q_{\rm blue} (p,x) \;,
 \\ [2mm]  
\bar q_8(p,x) &=&  \frac{1}{\sqrt{3}} \Big( \bar q_{\rm red} (p,x) 
     +  \bar q_{\rm blue}(p,x) - 2\bar  q_{\rm green} (p,x)  \Big)  \;
\ea
\end{mathletters}
where the distribution functions of quarks and antiquarks of different 
colors are of the form (\ref{qlocalup}) and (\ref{barlocalup}), respectively, 
but with the following color chemical potentials:
\be
\mu_{\rm red}(x)  = \frac 12 \left( \mu_3 (x) 
+ \frac{\mu_8 (x)}{\sqrt{3}} \right)  
\;,  \qquad
\mu_{\rm blue}(x)  =   - \frac 12 \left( \mu_3 (x) - 
\frac{\mu_8 (x)}{\sqrt{3}} \right) 
\;, \qquad
\mu_{\rm green}(x)  =   - \frac{ \mu_8 (x)}{\sqrt{3}} \;. 
\ee

The computation of  the singlet and adjoint components of the
local equilibrium distribution function of gluons is much more
involved. The evaluation of the traces requires to expand the
exponentials, and compute the traces of arbitrary powers of  
$T_3$, of $T_8$, and of $T_3 T_8$. With the help of Mathematica,
we have found the singlet and (non-vanishing) adjoint components as
\ba
g_0 (p,x) &=& 2 g_{\rm s} (p,x) + g_{\rm x+}(p,x) + g_{\rm x-}(p,x) 
+  g_{\rm y+}(p,x) + g_{\rm y-}(p,x) +g_{\rm z+}(p,x) + g_{\rm z-}(p,x) 
\;, \\
g_3(p,x) & = & g_{\rm z+}(p,x) - g_{\rm z-}(p,x) 
+ \frac 12 \Big( g_{\rm x+}(p,x) + g_{\rm x-}(p,x) 
+  g_{\rm y+}(p,x) + g_{\rm y-}(p,x)  \Big) 
\;, \\
g_8 (p,x) & = & {\sqrt{3} \over 2}\, 
\Big(g_{\rm x+}(p,x) - g_{\rm x-}(p,x) -
g_{\rm y+}(p,x) + g_{\rm y-}(p,x) \Big)  \;,
\ea
where the scalar functions $g_{\rm s}$, $g_{\rm x \pm}$, $g_{\rm y \pm}$,
and $g_{\rm z \pm}$ are analogous to those from Eqs.~(\ref{glocalup})
but their color chemical potentials are 
\be
\mu_{\rm s}(x) =  0 \;, \qquad
\mu_{\rm x \pm}(x) = \pm \frac{\mu_3(x)}{2} 
               \pm \frac{\sqrt{3} \mu_8(x)}{2} \;, \qquad
\mu_{\rm y \pm}(x) = \pm \frac{\mu_3(x)}{2} 
               \mp \frac{\sqrt{3} \mu_8(x)}{2} \;, \qquad 
\mu_{\rm z \pm}(x) = \pm \mu_3(x)  \;. 
\ee
Exactly as in the SU(2) case, we find that a finite value of
the color chemical potential means that quarks, antiquarks and
gluons of different colors have different densities.


\section{Local equilibrium from vanishing collision terms}
\label{loc-col}


As follows from Eq.~(\ref{entro-prod}), there is no entropy production
when the collision terms vanish. Thus,  local equilibrium is reached
when the gain and loss terms compensate each other. Consequently, we will 
look for solutions of the equation $C = 0$. However, there are numerous 
scattering processes occurring in the quark-gluon plasma and, in general, 
the complete set of collision terms entering into the quark, antiquark 
and gluon kinetic equations is rather large, even so we only consider 
the binary collisions. The most probable processes, {\it i.e.} those 
with the largest cross section, occur when two partons exchange a soft 
gluon in the $t$- or $u$-channels. The later possibility only happens 
for interaction of identical partons - quarks of the same flavour or
gluons.  In vacuum, the corresponding cross sections diverge as $t^{-2}$ or
$u^{-2}$ when the four-momentum transfer $t$ or $u$ goes to zero. In the 
medium, these divergences are softened, as the gluon propagators are dressed 
by the interactions, and the electric and magnetic forces are either 
statically or dynamically screened. In the local equilibrium state, which 
is achieved at the shortest time scale, the collision terms associated with 
those processes, we call them `dominant', have to vanish. Thus, we will 
first consider the interactions: 
$qq \leftrightarrow qq$, $\bar q \bar q  \leftrightarrow \bar q \bar q$,
$q \bar q \leftrightarrow q \bar q$,
$gg \leftrightarrow gg$, $qg \leftrightarrow qg$, and
$\bar qg \leftrightarrow \bar q g$, and we  will neglect all other processes,
as they are relevant for longer time scales\footnote{To estimate a mean
free time associated with a given collision process one has to specify
the distribution function. For a discussion of those mean free times in
global equilibrium see Ref. \cite{Arnold:2002zm}.}. These less probable
processes drive the system either to a different local equilibrium, or
to the global equilibrium. We will also consider the subdominant
processes with the soft quark in $t$- or $u$-channel which correspond
to the vacuum cross sections diverging as $t^{-1}$ or $u^{-1}$, respectively.
These are the quark-antiquark annihilation and creation into and from
two gluons in $t$- or $u$-channel which, as will be shown, have a
qualitative effect on the local equilibrium state. With the subdominant
processes, one should also consider all the channels and the respective
crossing terms of the various binary collisions, plus another set of
collisions that do not conserve the particle number. The complete analysis 
is very complex, and we will not carry it out here.

In this section we write down the relevant collision terms, and then we 
discuss the equations imposed by the vanishing of these terms. Finally, we 
solve the equations, showing that the nature of local equilibrium is 
fixed by the color structure of the scattering amplitudes.


\subsection{Collision terms}
\label{coll-term}


The dominant parton-parton scattering amplitudes with one-gluon exchange in the 
$t$- and $u$-channels are of the form
\ba
\label{sct-tchannel}
M_{rsr's'}(p,p_1;p',p'_1) &=& {\cal M}(p,p_1;p',p'_1) \; {\cal T}^a_{rr'}
 \tilde {\cal T}^a_{ss'} \\[2mm]
\label{sct-uchannel}
M_{rsr's'}(p,p_1;p',p'_1) &=& {\cal M}(p,p_1;p',p'_1) \; {\cal T}^a_{rs'}
 \tilde {\cal T}^a_{sr'} \;,
\ea
where ${\cal T}^a$ and $\tilde {\cal T}^a$ are the group generators
of SU($N_c$) of the two partons participating in the collision:
${\cal T}^a = T^a$ for gluons, ${\cal T}^a = \tau^a$ for quarks
and ${\cal T}^a = - (\tau^a)^T$, where $T$ means transposition,
for antiquarks. With the $t$-channel amplitude (\ref{sct-tchannel}), 
the collision term (\ref{Wald}) equals
\ba \label{Wald-t}
C[N,N_1,N',N_1']  &=&  \int dP' dP_1'dP_1 \;
(2 \pi)^4 \delta^{(4)} (p+p_1 -p'-p_1') \; |{\cal M}|^2 \\ [2mm]
\nonumber &\times&
\Big({\cal T}^a N' {\cal T}^b \;
{\rm Tr}[\tilde {\cal T}^a N_1' \tilde {\cal T}^b]
- \frac{1}{2} \{{\cal T}^b {\cal T}^a , N \}
\; {\rm Tr}[\tilde {\cal T}^a N_1 \tilde   {\cal T}^b] \Big) \;,
\ea
where we have neglected the effects of quantum statistics, and
consequently the terms $1\pm N$ have been replaced by unity. 
The collision term corresponding to the $u$-channel amplitude 
(\ref{sct-uchannel}) can be found from (\ref{Wald-t}) by means 
of the exchange $N \leftrightarrow N_1$ and $N' \leftrightarrow N'_1$ 
in the r.h.s of Eq.~(\ref{Wald-t}).

Using the identity (\ref{id-fund}) given in the Appendix,
we can write down Eq.~(\ref{Wald-t}) for the case of quark-quark
scattering as
\ba \label{Wald-qq-t}
C[Q,Q_1,Q',Q_1']  &=&  {1 \over 2} \int dP' dP_1'dP_1 \;
(2 \pi)^4 \delta^{(4)} (p+p_1 -p'-p_1') \; |{\cal M}|^2 \\ [2mm]
\nonumber &\times&
\Big(\big({\rm Tr}[Q']Q_1' - {\rm Tr}[Q]Q_1 \big)
- \frac{1}{N_c^2} \big(Q'{\rm Tr}[Q_1'] - Q{\rm Tr}[Q_1] \big)
- \frac{1}{N_c} \big(\{Q',Q_1'\} - \{Q,Q_1\} \big) \Big) \;.
\ea
The collision term (\ref{Wald-t}) for the quark-antiquark scattering is
\ba \label{Wald-qaq-t}
C[Q,\bar Q_1,Q',\bar Q_1']  &=&  {1 \over 2} \int dP' dP_1'dP_1 \;
(2 \pi)^4 \delta^{(4)} (p+p_1 -p'-p_1') \; |{\cal M}|^2 \\ [2mm]
\nonumber &\times&
\Big( \big( {\rm Tr}[Q'\bar Q_1'] - \frac{N_c}{2} \{Q, \bar Q_1\} \big)
- \frac{1}{N_c^2} \big(Q'{\rm Tr}[\bar Q_1'] - Q{\rm Tr}[ \bar Q_1] \big)
- \frac{1}{N_c} \big(\{Q',\bar Q_1'\} - \{Q,\bar Q_1\} \big) \Big) \;,
\ea
where, as discussed previously, we have replaced $\bar Q^T$ by $\bar Q$.

For the gluon-gluon scattering we have found a simplification of 
Eq.~(\ref{Wald-t}) only in the case of the SU(2) gauge group. Then,
the collision term reads
\ba \label{Wald-gg-t}
C[G,G_1,G',G_1']  &=&  \int dP' dP_1'dP_1 \;
(2 \pi)^4 \delta^{(4)} (p+p_1 -p'-p_1') \; |{\cal M}|^2 \\ [2mm]
\nonumber &\times&
\Big(\big({\rm Tr}[G'^T G'_1] - \{G'^T, G'_1 \}
- \frac{1}{2} \{G, G^T_1 \} \big)
+ \big({\rm Tr}[G'] \; G'_1 - G \; {\rm Tr}[G_1] \big) \Big) \;.
\ea

The scattering amplitudes of the subdominant processes with the quark 
exchange in $t$- and $u$-channel have the following color structure
\ba
\label{creat-tchannel}
M_{ijab}(p,p_1;p',p'_1) & = & {\cal M}(p,p_1;p',p'_1) \tau^a_{ik}
\tau^b_{kj} \;, \\
\label{creat-uchannel}
M_{ijab}(p,p_1;p',p'_1) & = & {\cal M}(p,p_1;p',p'_1) \tau^b_{ik}
\tau^a_{kj} \;. 
\ea
The collision term associated with this $t$-channel annihilation
processes is 
\ba \label{Wald-t-annihi}
C[Q,\bar Q_1,G',G_1']  &=&  \int dP' dP_1'dP_1 \;
(2 \pi)^4 \delta^{(4)} (p+p_1 -p'-p_1') \; |{\cal M}|^2  \\ [2mm]
\nonumber &\times&
\Big(\tau^a \tau^b \tau^{\bar b} \tau^{\bar a} G^{a \bar a} (p') 
G^{b \bar b} (p'_1) - 
\frac{1}{2} \{Q(p), \tau^a \tau^b \bar Q (p_1)\tau^b \tau^a \}\Big) 
\;.
\ea

At the end of this section we call the attention of the reader to the
structure of the collision terms (\ref{Wald-qq-t},\ref{Wald-qaq-t},
\ref{Wald-gg-t}). Because there are only objects like $Q$, $G$, $G^T$,  
which transform covariantly with respect to the gauge transformation 
(\ref{Q-transform},\ref{G-transform}), and ${\rm Tr}[Q]$, ${\rm Tr}[G]$ 
and ${\rm Tr}[Q \bar Q_1]$, which are gauge invariant, these collision 
terms transform covariantly, provided $|{\cal M}|^2$ is gauge invariant. 
The gauge covariance of the collision term (\ref{Wald-t-annihi}) is 
evident when instead of $C$ the projections ${\rm Tr}[C]$ and 
${\rm Tr}[\tau^aC]$ are considered. As will be seen in the following 
subsection, these projections have the right gauge structure.


\subsection{Conditions of local equilibrium}


In this subsection we present the conditions for the cancellation 
of the collision terms associated with the processes discussed above.

\subsubsection{$qq \leftrightarrow qq$}

The collision term (\ref{Wald-qq-t}) corresponding to the quark-quark 
scattering vanishes if 
\be \label{cancel-Wald-qq-t}
\big({\rm Tr}[Q']Q_1' - {\rm Tr}[Q]Q_1 \big)
- \frac{1}{N_c^2} \big(Q'{\rm Tr}[Q_1'] - Q{\rm Tr}[Q_1] \big)
- \frac{1}{N_c} \big(\{Q',Q_1'\} - \{Q,Q_1\} \big) = 0 \;,
\ee
where $ p + p_1 = p'+ p'_1$. Because the quark matrix transport equation
can be uniquely characterized by its singlet and adjoint components 
(see Eqs.~(\ref{projec-transport})), the condition (\ref{cancel-Wald-qq-t}) 
requires
\begin{mathletters}
\label{cancel-proj-qq}
\ba
\label{qq-a}
{\rm Tr} [Q Q_1] & = & {\rm Tr} [Q' Q'_1] \;, \\
\label{qq-b}
{\rm Tr} [Q] \; {\rm Tr}[Q_1] & = & {\rm Tr} [Q'] \; {\rm Tr} [Q'_1] \;,
\ea
and
\ba
\label{qq-c}
{\rm Tr} [\tau^a \left\{Q, Q_1\right\}] & = &
 {\rm Tr} [\tau^a \left\{Q', Q'_1\right\}]  \;,\\
\label{qq-d}
{\rm Tr} [\tau^a Q]\; {\rm Tr}[Q_1] & = & {\rm Tr} [Q'] \;
{\rm Tr} [\tau^a Q'_1]  \;, \\
\label{qq-e}
{\rm Tr} [\tau^a Q] \; {\rm Tr}[Q_1] & = &
 {\rm Tr} [\tau^a Q'] \;
{\rm Tr} [Q'_1]  \;.
\ea
\end{mathletters}

The conditions for cancellation of the collision term for antiquark-antiquark 
scattering are totally analogous to those of the quark-quark case.

\subsubsection{$q \bar q \leftrightarrow q \bar q$}

The collision term (\ref{Wald-t}) for the quark-antiquark scattering 
vanishes when
\be \label{cancel-Wald-qaq-t}
\big( {\rm Tr}[Q'\bar Q_1'] - \frac{N_c}{2} \{Q, \bar Q_1\} \big)
- \frac{1}{N_c^2} \big(Q'{\rm Tr}[\bar Q_1'] - Q{\rm Tr}[ \bar Q_1] \big)
- \frac{1}{N_c} \big(\{Q',\bar Q_1'\} - \{Q,\bar Q_1\} \big) = 0 \;.
\ee
The conditions of cancellation of the projected matrix equation 
(\ref{cancel-Wald-qaq-t}) read
\begin{mathletters}
\label{cancel-proj-qbarq}
\ba
\label{qbarq-a}
{\rm Tr} [Q \bar Q_1] & = & {\rm Tr} [Q' \bar Q'_1]  \;, \\
\label{qbarq-b}
{\rm Tr} [Q] \; {\rm Tr}[\bar Q_1] & = & {\rm Tr} [Q'] \;
{\rm Tr} [\bar Q'_1]  \;,
\ea
and
\ba
\label{qbarq-c}
{\rm Tr} [\tau^a \left\{Q, \bar Q_1\right\}] & = &
 {\rm Tr} [\tau^a \left\{Q', \bar Q'_1\right\}] = 0  \;,\\
\label{qbarq-d}
{\rm Tr} [\tau^a Q]\; {\rm Tr}[\bar Q_1] & = &
{\rm Tr}[ \tau^a Q'] \;
{\rm Tr} [\bar Q'_1]  \;.
\ea
\end{mathletters}
The requirement that ${\rm Tr} [\tau^a \left\{Q, \bar Q_1\right\}] = 0$
directly follows from the first term of Eq.~(\ref{cancel-Wald-qaq-t}). 

\subsubsection{$qg \leftrightarrow qg$}

The collision term for quark-gluon scattering with one-gluon exchange in the 
$t$-channel vanishes if
\be \label{cancel-Wald-qg-t}
\tau^a Q' \tau^b \; {\rm Tr}[T^a G_1' T^b]
- \frac{1}{2} \{\tau^b \tau^a , Q \} 
\; {\rm Tr}[T^a G_1 T^b] = 0\;.
\ee
Requiring that ${\rm Tr}[C] = 0$ and ${\rm Tr}[\tau^a C] = 0$
provides the equations
\begin{mathletters}
\ba
\label{qg-a}
&{\rm Tr} [Q] \; {\rm Tr}[G_1] - \frac{N_c}{2}
 {\rm Tr} [\tau^a Q] \; {\rm Tr}[ T^a G_1]
+ \frac 12 d^{abc}\, {\rm Tr} [\tau^a Q] \; {\rm Tr}[ T^b T^c G_1]&
\nonumber \\
 =
&{\rm Tr} [Q'] \; {\rm Tr}[G'] - \frac{N_c}{2}
 {\rm Tr} [\tau^a Q'] \; {\rm Tr}[ T^a G'_1]
+ \frac 12 d^{abc} \,{\rm Tr} [\tau^a Q'] \; {\rm Tr}[ T^b T^c G'_1]&  \;,
\ea
and
\be
\label{qg-b}
{\rm Tr} [\tau^c \tau^b Q \tau^a] \; {\rm Tr}[ T^b G_1 T^a]
+ R^c [Q,G_1] 
= {\rm Tr} [\tau^c \tau^b Q' \tau^a] \; {\rm Tr}[ T^b G'_1 T^a] \;,
\ee
where
\be
\label{qg-c}
 R^c [Q,G_1] \equiv \frac i2 f_{c a d} \big( 
{\rm Tr} [\tau^d  \tau^b Q ] \; {\rm Tr}[ T^b G_1 T^a]
- {\rm Tr} [\tau^b  \tau^d Q ] \; {\rm Tr}[ T^a G_1 T^b]\big) \;.
\ee
\end{mathletters}

\subsubsection{$gg \leftrightarrow gg$}

The collision term of the gluon-gluon scattering equals zero when

\be
\label{cancel-gg}
 T^a G'  T^b \;
{\rm Tr}[  T^a G_1'   T^b]
- \frac{1}{2} \{ T^b  T^a , G \}
\; {\rm Tr}[ T^a G_1     T^b]  = 0 \;.
\ee
For the SU(2) plasma the above condition can be simplified
(see Eq.~(\ref{Wald-gg-t})) and it gives
\be 
\label{cancel-Wald-gg-t}
\big({\rm Tr}[G'^T G'_1] - \{G'^T, G'_1 \} 
- \frac{1}{2} \{G, G^T_1 \} \big)
+ \big({\rm Tr}[G'] \; G'_1 - G \; {\rm Tr}[G_1] \big) = 0  \;.
\ee
We demand the cancellation of the totally symmetric and antisymmetric 
components of (\ref{cancel-Wald-gg-t}), see Eqs.(\ref{projec-transport-gSU2}).
Imposing ${\rm Tr}[T_a C_g] = 0$ and ${\rm Tr}[\{T_a,T_b\} C_g] = 0$,
we get
\begin{mathletters}
\ba
\label{gg-c}
{\rm Tr} [T_a \left\{G, G^T_1\right\}] & = &
 {\rm Tr} [T_a \left\{G', G'^T_1\right\}] = 0  \;,\\
\label{gg-d}
{\rm Tr} [T_a G]\, {\rm Tr}[G_1] & = & {\rm Tr} [G'] \,
{\rm Tr} [T_a G'_1]  \;,
\ea
and
\ba
\label{gg-pe}
{\rm Tr} [\{T_a,T_b\} \left\{G, G^T_1\right\}] & = &
8\delta^{ab} \: {\rm Tr} [G' G'^T_1]
- 2 {\rm Tr} [\{T_a,T_b \} \left\{G', G'^T_1\right\}]  \\
\label{gg-f}
{\rm Tr} [\{T_a,T_b \} G]\; {\rm Tr}[G_1] & = & {\rm Tr} [G'] \;
{\rm Tr} [\{T_a,T_b \} G'_1]  \;.
\ea
For $a \neq b$ Eq.~(\ref{gg-pe}) requires that
\be
\label{gg-e}
{\rm Tr} [\{T_a,T_b\} \left\{G, G^T_1\right\}]  = 
 {\rm Tr} [\{T_a,T_b \} \left\{G', G'^T_1\right\}] = 0 \;,
\ee
while for $a = b$ ($T^aT^a = 2$) Eqs.~(\ref{gg-pe},\ref{gg-f}) imply
\ba
\label{gg-a}
{\rm Tr} [G G^T_1] & = & {\rm Tr} [G' G'^T_1] \;, \\
\label{gg-b}
{\rm Tr} [G] \; {\rm Tr}[G_1] & = & {\rm Tr} [G'] \;
{\rm Tr} [G'_1]  \;.
\ea
\end{mathletters}

\subsubsection{$q \bar q \leftrightarrow gg$}

With the scattering amplitude given in Eq. (\ref{creat-tchannel}), 
the cancellation of the collision term corresponding to the 
quark-antiquark annihilation in the $t$-channel demands
\begin{mathletters}
\label{qbarqgg-t}
\ba
\label{qqgg-at}
{\rm Tr} [\tau^a Q \tau^a  \tau^b \bar Q_1 \tau^b ] &  = &
 {\rm Tr} [\tau^c \tau^d \tau^{\bar d} \tau^{\bar c}] \; 
G'^{c \bar c} G'^{d \bar d}_1 \;, \\
\label{qqgg-bt}
\frac 12
{\rm Tr} [\tau^e \left\{ Q, \tau^a \tau^b  \bar Q_1 \tau^b \tau^a
 \right\}] &  = &
{\rm Tr} [\tau^e \tau^c \tau^d \tau^{\bar d} \tau^{\bar c}] \; 
G'^{c \bar c} G'^{d \bar d}_1 \;.
\ea
\end{mathletters}
The left hand side of the above equations can be simplified
using the relations (\ref{id-fund-secun}) given in the
Appendix and the formula
\be
\tau^a \tau^b  \bar Q_1 \tau^b \tau^a 
= \frac{1}{4 N^2_c} \bar Q_1 + \frac{N_c^2 -2}{4 N_c} {\rm Tr}[\bar Q_1] \;,
\ee
Furthermore, for the SU(2) plasma one finds, using the relations 
(\ref{4-tau}, \ref{5-tau}) given in the Appendix, that
\begin{mathletters}
\label{SU2qbarqgg-t}
\ba
\label{SU2qqgg-at}
{\rm Tr}[Q \bar Q_1] + 4 {\rm Tr}[ Q] \; {\rm Tr}[\bar Q_1] 
&  = &  
2 {\rm Tr}[G'^T  G'_1] - 2 {\rm Tr}[G'  G'_1] 
+ 2 {\rm Tr}[G'] \; {\rm Tr}[G'_1] \;,
\\ \label{SU2qqgg-bt}
{\rm Tr}[\tau^e \left\{Q ,\bar Q_1 \right\}] 
+ 8 {\rm Tr}[\tau^e Q] \; {\rm Tr}[\bar Q_1]
& = & 2 G'^{ce} {\rm Tr}[T^c G'_1] + 2 {\rm Tr}[T^e G'^T_1  G']
- 2 {\rm Tr}[T^e G'_1  G'] + 2 {\rm Tr}[T^e G'] \; {\rm Tr}[G'_1] \;.
\ea
\end{mathletters}


\subsection{Local Equilibrium Solution for the SU(2) plasma}
\label{loc-eq-sol}


We find here the local equilibrium solutions that cancel all the collision
terms discussed in the previous subsection for the SU(2) plasma.
We start with the quark-quark 
scattering. Eqs. (\ref{qq-a},\ref{qq-c}) are solved by  functions 
obeying
\be \label{eq-qqqq}
 Q(p,x) \: Q(p_1,x) = Q(p',x) \: Q(p'_1,x)  \;,
\ee
for $p + p_1 = p' + p'_1$. Using  standard arguments, see {\it e.g.} 
\cite{Gro80}, one finds that Eq.~(\ref{eq-qqqq}) is satisfied by 
exponential functions
\be
 Q(p,x) = {\rm exp}\Big[- \beta (x) \big({\tilde u}_\nu (x) p^\nu 
- \mu_b(x) - \widetilde \mu (x) \big)\Big]   \;,
\ee
where ${\tilde u}^\mu(x)$ and $ {\tilde \mu}(x)$ are hermitian and 
traceless matrices. Please note that the scalar chemical potential
$\mu_b$, which is interpreted as the baryon chemical potential, is already
singled out. Because of Eq.~(\ref{diagonalcondition}), ${\tilde u}^\mu(x)$
and $ {\tilde \mu}(x)$ should obey the condition 
$[{\tilde u}^\mu(x),{\tilde \mu}(x)]= 0$. Thus, using the gauge freedom
to rotate these quantities in color space, they can be chosen in  
diagonal form. 

Eqs.~(\ref{qq-b},\ref{qq-d},\ref{qq-e}) require that the hydrodynamic
velocity ${\tilde u}^\mu(x)$ is proportional to the unit matrix.
Otherwise different components of ${\tilde u}^\mu(x)$ enter 
differently Eqs.~(\ref{qq-b},\ref{qq-d},\ref{qq-e}) and the
constraint $p + p_1 = p' + p'_1$ is insufficient to satisfy these
equations. Once ${\tilde u}^\mu(x)$ is proportional to the unit 
matrix, the condition $[{\tilde u}^\mu(x),{\tilde \mu}(x)] = 0$
is trivially satisfied, and there is no reason to require 
${\tilde \mu}(x)$ to be diagonal. It is then an arbitrary traceless
matrix, even so it can be diagonalized because of the gauge freedom.
Since the hydrodynamic velocity is no longer a color matrix but a scalar, 
it is from now on denoted as $u^\mu$ not as ${\tilde u}^\mu$. 

Repeating fully analogous considerations for the collision term
of antiquark-antiquark scattering, we arrive to the antiquark
distribution function
\be
\bar Q(p,x) = {\rm exp}\Big[-
\bar \beta (x) \big({\bar u}_\nu (x) p^\nu + \bar \mu_b(x) 
+ \widetilde {\bar \mu} (x) \big)\Big] \;.
 \ee

The conditions of cancellation for the quark-antiquark collision term
provide the relations between the parameters of quark and antiquark
distribution functions. Namely, 
Eqs.~(\ref{qbarq-a},\ref{qbarq-b},\ref{qbarq-d}) require
\be
\beta(x) \: u^\mu (x) = \bar \beta(x) \: \bar u^\mu (x) \;.
\ee
Because $u^\mu (x) u_\mu (x) = \bar u^\mu (x) \bar u_\mu (x) = 1$,
we effectively have 
\be
u^\mu (x) = \bar u^\mu (x) \;, \qquad T(x) = \bar T(x) \ .
\ee
Furthermore, Eq.~(\ref{qbarq-c}) imposes
\be
\widetilde{\mu} (x) = - \widetilde{\bar\mu} (x)  \; , 
\ee
but it leaves the baryon chemical potentials $\mu_b$ and $\bar \mu_b$ 
unrestricted.

Let us find now the distribution functions that cancel the gluon-gluon
collision term. Condition (\ref{gg-a}) is solved by those functions
obeying
\be
\label{prod-gg}
G(p,x) G^T(p_1,x)  = G(p',x) G^T(p'_1,x)  \;,
\ee
which demands that
\be
\label{local-gSU2}
G(p,x) = {\rm exp}\Big[-\beta_g (x)
\big({\widetilde u}_g^ \nu (x) p_\nu -  \mu_g^0(x) 
-\widetilde \mu_g (x) \big)\Big]
\ee
where both ${\widetilde u}_g^\mu(x)$ and ${\widetilde \mu}_g(x)$ are 
hermitian matrices while $ \mu_g^0(x)$ is a scalar. Furthermore, 
${\tilde u}_g^\mu(x)= ({\tilde u}_g^\mu(x))^T$, which implies that
${\tilde u}_g^\mu(x)$ is a real symmetric matrix. However, the conditions 
(\ref{gg-b}) and (\ref{gg-d}) require that the gluon velocity matrix has 
to be proportional to the unit matrix, exactly as that of quarks
and antiquarks.

The condition (\ref{gg-c}) or (\ref{gg-e}) implies that
the product $G G_1^T$ must be proportional to the unit matrix. 
Therefore, the gluon color chemical potential must obey 
${\widetilde \mu_g}^T (x) = - \widetilde \mu_g (x)$. Consequently,
it contains only antisymmetric components and it can be uniquely 
expressed as $\widetilde \mu_g (x) = \mu^a_g (x) T^a$. 

Next, we analyze the conditions for cancellation of the quark-gluon 
collision term {\it i.e.} Eqs.~({\ref{qg-a},\ref{qg-b}). For SU(2)
$d_{abc} = 0$, and then it is easy to check that  Eq.~({\ref{qg-a}) 
imposes
\be
 T(x) = T_g (x) \ , \qquad u^\mu (x) =  u^\mu_g (x) \;.
\ee
Thus, the temperature, as well as the hydrodynamic velocity, are the
same for the quark-antiquark and gluon components of the plasma. 

Eq.~({\ref{qg-b}) is of more complicated structure. Since it is fulfilled 
if $R^c = 0$, let us evaluate $R^c$. Taking into account that for SU(2)
\be
{\rm Tr}[\tau^a  \tau^b Q(p,x)] = \frac i4 f^{abc} \, q_c (p,x) 
+ \frac 14 \delta^{ab} \, q_0 (p,x) \:,
\ee
and using the relation (\ref{id-ad-2}) given in the Appendix, 
we express $R^c$ as
\be
R^c = -\frac 18 q_a (p,x) \; 
{\rm Tr}[\left\{T^a,T^c \right\} G(p_1,x)]
+ \frac 12 q_c(p,x) \; g_0 (p_1,x) 
- \frac 14 q_0(p,x) \; g_c (p_1,x) \ .
\ee
And now we refer to the diagonal gauge where the quark chemical
potential is of the form $\widetilde \mu (x) = \mu_3 (x) \tau^3$. 
Requiring $R^1=R^2 = 0$ implies $g_1 = g_2= 0$, which, in turn, demands
that the respective components of the gluon chemical potential
vanish {\it i.e.} $\mu^1_g  = \mu^2_g = 0$. Demanding $R^3 = 0$ is 
only fulfilled if 
\be
\mu_3 (x) = \mu^3_g (x)  \;.
\ee
Thus, Eq.~({\ref{qg-b}) is satisfied in arbitrary gauge if the relation
(\ref{relat-qgmu}) holds.  

The dominant processes that have been considered till now  do not
introduce any relation between the quark and antiquark baryon chemical
potentials and they do not constrain the scalar gluon potential $\mu_g^0$. 
It is not surprising as these processes do not change the number of 
quarks, antiquarks or gluons. To get the relation between $\mu_b$,
$\bar \mu_b$ and $\mu_g^0$, the subdominant process of quark-antiquark 
creation or annihilation has to be taken into account. Let us analyze this 
process.  The color structure of Eqs.~(\ref{SU2qqgg-at},\ref{SU2qqgg-bt})
is rather complex. However, one checks that these equations
are solved by the local equilibrium function (\ref{loc-eq})
in the diagonal gauge (\ref{loc-eq-q-diag-SU2},\ref{loc-eq-g-diag-SU2}).
In particular, one finds that
\ba
{\rm Tr}[Q \bar Q_1] + 4 {\rm Tr}[ Q] \; {\rm Tr}[\bar Q_1] 
&  = & e^{- \beta\left( u \cdot (p+p_1) -\mu_b +
\bar \mu_b  \right) } \Big(10 + 4\; e^{ \beta \mu_3}
+ 4\; e^{- \beta \mu_3} \Big) \;,
\\
2 {\rm Tr}[G'^T  G'_1]  -  2 {\rm Tr}[G'  G'_1] 
& + & 2 {\rm Tr}[G'] \; {\rm Tr}[G'_1] = 
e^{- \beta\left( u \cdot (p'+p'_1) - 2\mu^0_g 
 \right) } \Big(10 + 4 \;e^{ \beta \mu_3}
+ 4 \; e^{- \beta \mu_3} \Big) \;.
\ea
Thus,   Eq.~(\ref{SU2qqgg-at}) demands
\be \label{rel-mub-mug}
\mu_b + \bar \mu_b = 2 \mu_g^0 \;.
\ee
While the checking is rather simple for Eq.~(\ref{SU2qqgg-at}), 
it is much more difficult for Eq.~(\ref{SU2qqgg-bt}). To reach
the goal we have expressed the (anti-)quark and gluon distribution 
functions through the projections (\ref{quprojec-component})
and (\ref{gluonbasis2}), respectively, and we have used the formula 
(\ref{G-vs-g}). Then, one finds that Eq.~(\ref{SU2qqgg-bt}) is 
satisfied if the relation (\ref{rel-mub-mug}) holds.

To get the chemical potentials as in the local equilibrium function
(\ref{loc-eq}) the binary processes are insufficient. One easily
observes that the equilibrium with respect to the process
$gg \leftrightarrow ggg$ implies $\mu_g^0 = 0$. And then,
Eq.~(\ref{rel-mub-mug}) provides $ \mu_b = - \bar \mu_b$.

In summary, the requirement of equilibrium with respect to the dominant 
binary processes provides the local equilibrium state with the color 
structure as that in (\ref{loc-eq}) which comes from the collisional
invariants. The (scalar) chemical potentials of quarks, antiquarks
and gluons are, however, independent from each other. To get the 
relations $\mu_g^0 = 0$ and $\mu_b = - \bar \mu_b$, the multigluon 
processes and antiquark-quark annihilation into gluons must be
taken into account. This means that the local {\em chemical} equilibrium 
is reached at longer time scale than the color equilibrium.


\subsection{Local Equilibrium Solution for the SU($N_c$) plasma}
\label{loc-eq-solSUN}


We find here the local equilibrium solutions for the SU($N_c$)  plasma.
The quark-quark and antiquark-quark scattering processes are treated
as for the SU(2) case. The solutions of 
Eqs.~(\ref{cancel-proj-qq},\ref{cancel-proj-qbarq})  read
\ba
 Q(p,x) & = & {\rm exp}\Big[- \beta (x) \big({\tilde u}_\nu (x) p^\nu 
- \mu_b(x) - \widetilde \mu (x) \big)\Big]   \;, \\
\bar Q(p,x) & = & {\rm exp}\Big[-
 \beta (x) \big( u_\nu (x) p^\nu + \bar \mu_b(x) 
+ \widetilde  \mu (x) \big)\Big] \;.
\ea

The conditions for cancellation of the collision terms discussed in 
Sec.~\ref{coll-term}, which involve gluons, are much more complicated 
than those for SU(2). Here, we will treat them perturbatively only.
The requirement of vanishing of the collision term representing gluon-gluon 
scattering is expressed by Eq.~(\ref{cancel-gg}). We first note that a 
distribution function proportional to the identity matrix, which is of 
the form ${\rm exp}\big[-\beta_g \big(u_g^ \nu p_\nu 
-  \mu_g^0 \big)\big]$ satifies this equation. We now look 
for more general solutions written as
\be
\label{ans-geq}
G(p,x) = {\rm exp}\Big[-\beta_g (x)
\big(u_g^ \nu (x) p_\nu -  \mu_g^0(x)  \big)\Big] 
F[\widetilde \alpha (x)] \;,
\ee
where we have factored out the U(1) part of the distribution function; 
$F$ is an arbitrary function of 
$\widetilde \alpha (x) = \beta_g (x) \widetilde \mu_g (x)$ with
$\widetilde \mu_g (x)$ being any hermitian $(N_c^2 -1) \times (N_c^2 -1)$ 
matrix. From Eq.~(\ref{cancel-gg}) one deduces that $F$ should obey the 
quadratic equation
\be
\label{quadr-gg-f}
\Big(T^a F[\widetilde \alpha ] T^b 
- \frac 12 \left\{T^b T^a, F[\widetilde \alpha ]
 \right\} \Big) {\rm Tr} [F[\widetilde \alpha ] T^b T^a]  = 0 \;,
\ee 
which is trivially satisfied by the unit matrix. We now assume that $F$ 
allows for an infinitesimal expansion in $\widetilde \alpha$ around the identity.
Then, 
\be
F[\widetilde \alpha] = 1 + \widetilde \alpha + \cdots  \;,
\ee
and Eq.~(\ref{quadr-gg-f}) imposes 
\be
T^a [\widetilde \alpha, T^a ] 
+ \frac 12 T^c {\rm Tr}[T^c \widetilde \alpha] = 0 \;.
\ee
If $\widetilde \alpha$ is proportional to the unit matrix, the equation
is obviously satisfied. However, we exclude this possibility since a scalar 
chemical potential was already included in the U(1) part of 
Eq.~(\ref{ans-geq}). A different solution of the equation is given by 
$\widetilde \alpha = \alpha_a T^a$. With the last option, we solve 
Eq.~(\ref{quadr-gg-f}) to second order in $\widetilde \alpha $,
and find 
\be
F[\widetilde \alpha] = 1 + \widetilde \alpha 
+ \frac{\widetilde \alpha^2}{2} + \cdots
\ee

In principle, one can solve the equation iteratively order by order in 
$\widetilde \alpha$ but the procedure becomes more and more difficult
with every order. We will not follow it but the above results suggests
that the general solution is of the form
\be
G(p,x) = {\rm exp}\Big[-\beta_g (x)
\big(u_g^ \nu (x) p_\nu -  \mu_g^0(x) 
-\widetilde \mu_g (x) \big)\Big] \;,
\ee
as it should reduce to the $N_c =2$ solution (\ref{local-gSU2}) with
the scalar hydrodynamic velocity. 

We now look for the quark-gluon scattering, and solve 
Eq.~(\ref{cancel-Wald-qg-t}) perturbatively for small color chemical
potentials of quarks and gluons. In 0-th order, Eq.~(\ref{cancel-Wald-qg-t}) 
imposes 
\be
 T(x) = T_g (x) \ , \qquad u^\mu (x) =  u^\mu_g (x) \;.
\ee
In the first order in the color chemical potentials, we find that these
should obey 
\be
\Big(\tau^a \widetilde \mu(x) \tau^b 
- \frac 12 \left\{\tau^b \tau^a, \widetilde \mu (x) \right\}
\Big) {\rm Tr}[T^b T^a] +
[\tau^a,\tau^b] {\rm Tr}[\widetilde \mu_g (x) T^b T^a] = 0 \:,
\ee
which is only satisfied if
\be
\widetilde \mu_g (x) = 2 T^a {\rm Tr}[\widetilde \mu(x) \tau^a] \;.
\ee
One could go to higher orders in the expansion but the procedure
becomes very tedious. 

In the same way, one can treat the remaining processes such as 
the quark-antiquark annihilation. They lead to the same constraints 
as those for the SU(2) plasma expressed, in particular, by 
Eq.~(\ref{rel-mub-mug}).
 
The perturbative treatment presented here is concluded 
as follows. At zeroth order, the various collision processes allow one 
to fix the variables which are scalar in color space - the temperature 
and hydrodynamic velocity. At first order, every collision term imposes 
restrictions on the form of the matrix chemical potentials. Solving the 
conditions to all orders should simply provide the solutions, which for 
classical statistics are  exponential functions of color chemical 
potentials.


\section{Chromohydrodynamics}
\label{color-hydro}


The form of the local equilibrium distribution function determines
the character of hydrodynamics obeyed by the system. Here, we 
are going to discuss the hydrodynamic equations corresponding to 
the local equilibrium state found in the previous sections.
As we have shown in Sec. \ref{loc-eq-sol}, the dominant processes, 
which are responsible for establishing the colorful equilibrium, do 
not equilibrate the system with respect to the scalar chemical 
potentials. The relations $\mu_g^0 = 0$ and $ \mu_b = - \bar \mu_b$ 
are achieved at  longer time scales.  Since we are mostly 
interested here in the role of color charges in the hydrodynamic 
evolution, we neglect complications caused by the lack of chemical 
equilibrium and we use the distribution functions (\ref{loc-eq}) 
where the relations $\mu_g^0 = 0$ and $\mu_b = - \bar \mu_b$ are 
built in. 

The equations of hydrodynamics are provided by the macroscopic 
conservation laws of the baryon charge (\ref{bar-cons}), energy-momentum 
(\ref{en-mom-cons}) and of the color charge (\ref{col-cons}).
Substituting the local equilibrium distribution functions (\ref{loc-eq}) 
into Eqs.~(\ref{bar-current},\ref{en-mom},\ref{col-current}), one gets
the baryon current, the energy-momentum tensor and the color current
which enter the equations of {\em ideal} hydrodynamics where dissipative
effects are neglected. These quantities read
\begin{mathletters}
\label{ideal}
\ba \label{id-bar-flow}
b^{\mu}(x) &=&  b(x) \: u^{\mu}(x) \;,
\\ \label{id-en-mom}
t^{\mu \nu}(x) &=& \bigr[ \varepsilon (x)+ p(x) \bigl] \,
u^{\mu}(x) \: u^{\nu}(x)
-p(x) \; g^{\mu \nu} \;,
\\ \label{id-col-current}
j^{\mu}(x) &=&  \rho (x) \: u^{\mu}(x) \;,
\ea
\end{mathletters}
where $b$, $\varepsilon$ and $\rho$ are the densities of, respectively,
the baryon charge, energy and color, while $p$ denotes the pressure.
In contrast to $b$, $\varepsilon$ and $p$ which are color scalars,
the color density $\rho$ is a $N_c \times N_c$ matrix.  All these
thermodynamical quantities are given as
\ba
b &=& { 2 N_f T^3 \over 3 \pi^2} \;
\Big[ e^{\beta \mu_b} \: {\rm Tr}[e^{\beta \tilde \mu}]
- e^{-\beta \mu_b} \: {\rm Tr}[e^{-\beta \tilde \mu}] \Big] \;,
\\ [2mm]
\varepsilon &=& 3p = { 6T^4 \over \pi^2} \;
\Big[ N_f \big( e^{\beta \mu_b} \: {\rm Tr}[e^{\beta \tilde \mu}]
+ e^{-\beta \mu_b} \: {\rm Tr}[e^{-\beta \tilde \mu}]\Big)
+ {\rm Tr}[e^{\beta \tilde \mu_g}] \Big] \;,
\\ [2mm]
\rho &=& - g \,{T^3 \over \pi^2} \;
\Big[ N_f \Big( e^{\beta \mu_b} \big( e^{\beta \tilde \mu} -
{1 \over N_c}\, {\rm Tr}[e^{\beta \tilde \mu}] \big)
- e^{-\beta \mu_b} \big( e^{-\beta \tilde \mu} -
{1 \over N_c}\, {\rm Tr}[e^{-\beta \tilde \mu}]\big) \Big)
+ 2 \tau^a {\rm Tr}[T^a e^{\beta \tilde \mu_g}] \Big] \;.
\ea

Now, we consider Eq.~(\ref{en-mom-cons}) representing the energy-momentum
conservation. It is well known \cite{Lan63} that projecting the continuity 
equation of the energy-momentum tensor on the hydrodynamic velocity, 
one gets the condition of the entropy conservation during the fluid 
motion. Let us see how it works here. Multiplying Eq.~(\ref{en-mom-cons}) 
by $u^{\mu}$, we get
\be \label{isoentro1}
u_{\mu}\partial_{\nu} t^{\mu \nu} = 0 \;,
\ee
because $u_{\mu} u^{\mu} = 1$ and $u_{\mu} u_{\nu} F^{\mu \nu} = 0$. 
The latter equality holds due to the antisymmetry of $F^{\mu \nu}$. 
Eq.~(\ref{isoentro1}) gives
\be
u_\mu \partial^\mu \varepsilon
+ (\varepsilon + p) \, \partial^\mu u_\mu = 0 \;,
\ee
which can be rewritten as
\be \label{isoentro2}
T \partial_\mu (s u^\mu )
+ \mu_b \partial_\mu (b u^\mu )
+ {\rm Tr}[\widetilde \mu \, \partial_\mu ( \rho u^\mu )] = 0 \;,
\ee
by means of the thermodynamic relations
\ba
d\varepsilon & = & T ds + \mu_b db + {\rm Tr}[\widetilde \mu \, d\rho] \;,
\\
\varepsilon + p & = & Ts +  \mu_b b + {\rm Tr}[\widetilde \mu \, \rho] \;,
\ea
where $s$ is the (local) entropy density in the fluid rest frame.
The second term in Eq.~(\ref{isoentro2}) vanishes due to the 
conservation of the ideal baryon flow (\ref{id-bar-flow}) and
the third term also vanishes as
\be
{\rm Tr}[\widetilde \mu \, \partial_\mu ( \rho u^\mu )]
= {\rm Tr}[\widetilde \mu \, D_\mu ( \rho u^\mu )] = 0 \;.
\ee
The first equality holds because $\widetilde \mu$ and $\rho$ commute
with each other, and consequently 
${\rm Tr}\big[\widetilde \mu [A^\mu, \rho]\big] = 0$. The last equality 
expresses the covariant conservation of the ideal color current 
(\ref{id-col-current}). Thus, Eq.~(\ref{isoentro2}) finally gives 
the entropy conservation $\partial_\mu (s u^\mu ) = 0$.

The analog of the Euler's equation is obtained from Eq.~(\ref{en-mom-cons}),
projecting it onto direction perpendicular to $u^{\mu}$. Equivalently,
we consider the following combination of 
Eqs.~(\ref{en-mom-cons},\ref{isoentro1}) 
\be
\partial_{\nu} t^{\mu \nu}
- u^{\mu} u_{\nu}\partial_{\sigma} t^{\nu \sigma}
= 2{\rm Tr}[j_\sigma F^{ \nu \sigma} ]  \;,
\ee
which gives
\be \label{euler1}
(\varepsilon + p)\, u^{\nu} \partial_{\nu} u^{\mu} 
= \big(\partial^{\mu}- u^{\mu} u_{\nu}\partial^{\nu} \big) p 
+ 2{\rm Tr}[j_\nu F^{ \mu \nu } ]  \;.
\ee

To get a better insight of the physical meaning of Eq.~(\ref{euler1}) 
we write it down in the three-vector notation where
\be \label{3-vectors}
u^{\mu} \equiv (\gamma c, \gamma {\bf v}) \;, \qquad
j^{\mu}\equiv (c\rho , {\bf j})\;, \qquad
F^{0i} = E^i \;, \qquad F^{ij} = \epsilon_{ijk} B^k \;,
\ee
with $\gamma \equiv (1-{\bf v}^2/c^2)^{-1/2}$ and ${\bf E}$, ${\bf B}$ 
being the chromoelectric and chromomagnetic field, respectively. 
We have restored here the velocity of light $c$ to facilitate
taking the nonrelativistic limit of the derived hydrodynamic equation. 
Subtracting Eq.~(\ref{euler1}) for $\mu = 0$ multiplied by $v^i/c$ from 
Eq.~(\ref{euler1}) for $\mu = i$, one gets
\be
{\varepsilon + p \over 1 - {\bf v}^2/c^2} \; 
\Big( {\partial \over \partial t} + {\bf v} \nabla \Big) {\bf v}
= - \Big(\nabla + {1\over c^2}\, {\bf v} {\partial \over \partial t} \Big) p
- 2 {\rm Tr}[ \rho {\bf E} - {1\over c^2}\, {\bf v} ( {\bf j} \cdot {\bf E} )
+ {1\over c} \, {\bf j} \times {\bf B} ] \;.
\ee
which in the nonrelativistic domain $(v^2 \ll c^2)$ reads
\be \label{Euler-non-rel}
(\varepsilon + p) 
\Big( {\partial \over \partial t} + {\bf v} \nabla \Big) {\bf v}
= - \nabla p 
- 2 {\rm Tr}[ \rho {\bf E} + {1\over c} \, {\bf j} \times {\bf B} ] \;.
\ee
We note that the nonrelativistic limit, which is taken for the sake of
comparison with the analogous equation of the electron-ion plasma 
\cite{Kra73}, is only applied to the hydrodynamic velocity. The motion
of the fluid's constituents remains relativistic.

Although the quark-gluon plasma is composed of partons of several 
colors, the hydrodynamic equation (\ref{euler1}) describes a single
fluid. This happens because there is a unique hydrodynamic velocity
in the local equilibrium state (\ref{loc-eq}). Various color
components, which enter the energy-momentum tensor, do not neutralize
each other but they are `glued' together in the course of evolution. 
Such a single fluid chromohydrodynamics was briefly considered long 
ago \cite{Mrowczynski:1989bv} within  kinetic theory. An equation 
very similar to Eq.~(\ref{euler1}) has been 
recently derived \cite{Bistrovic:2002jx} directly from a postulated 
Lagrange density. The color current, which enters the Euler's equation
discussed in \cite{Bistrovic:2002jx}, is of the form 
$Q {\cal J}^{\mu}$ where $Q$ is the color charge and ${\cal J}^{\mu}$ 
is the conserved Abelian current. As seen in Eqs.~(\ref{ideal}),
${\cal J}^{\mu}$ can be identified with the baryon flow $b^{\mu}$ 
when we deal with a system of quarks only. In a multi-component plasma, 
however, such an identifications is not possible because vanishing
of the baryon current does not imply vanishing of the color current.


\section{Discussion and summary}
\label{discussion}


Local equilibrium is only a transient state of a non-equilibrium 
system in its course towards global equilibrium. Thus, the question
arises how fast such a state is achieved, and for how long it 
survives. We denote the two characteristic times of interest as 
$\tau_0$ and $\tau_1$. As we have shown in Sec.~\ref{loc-col}, the 
dominant processes, those with the soft gluon exchange in $t$ or $u$
channel, are responsible for establishing the colorful equilibrium.
Since the electric forces are screened at  momentum transfers 
smaller than the Debye mass ($m_D$) the largest contribution to these
processes  comes from the small angle scatterings due to the magnetic 
forces which are effectively screened at  momentum transfers below $m_D$.  
We identify $\tau_0$ with the relaxation time related to such 
interactions. Then, according to the estimate \cite{Arnold:1998cy}
found for the quark-gluon plasma in global (colorless) equilibrium
where $m_D \sim gT$, we have 
\be \label{time0}
{1 \over \tau_0} \sim g^2 T \; {\rm ln}(1/g) \;.
\ee
We note, however, that the relaxation time in the colorful plasma 
can significantly differ from (\ref{time0}) due to the interaction 
with the background chromodynamic field generated by the color current
(\ref{loc-cur}).

And for how long does the colorful equilibrium exist?
The answer crucially depends on the process which is responsible 
for the plasma neutralization. We have explicitly shown that the 
dominant processes comply with the finite color chemical potentials. 
We have also checked that equilibration with respect to the process 
$q \bar q \leftrightarrow gg$ leaves the system colorful. We expect 
that the collisions, even those beyond binary approximation, do not 
demand vanishing of the color chemical potentials. The point is that
in every collision process, which changes the particle momenta but 
not their `macroscopic' positions, the color current is conserved.
Therefore, the collisions do not alter the local color charge.

The plasma is presumably neutralized due to the collective phenomena: 
dissipative color currents and damp plasma waves both caused by 
uncompensated color charges. Then, the characteristic time of the 
system neutralization is controlled by the color conductivity which 
is again related to the estimate (\ref{time0}) \cite{Arnold:1998cy}. 
Thus, the two times of interest $\tau_0$ and $\tau_1$ are of the same 
order, and a much more careful analysis is needed to establish the 
domain of applicability of the local equilibrium solution found here. 
Such an analysis should take into account not only the interaction 
with the background fields present in the colorful equilibrium but 
the initial non-equilibrium configuration should be also specified. 

At the end let us summarize the most important results of this study.
The local equilibrium state dictated by the collisional invariants,
which follow from the energy-momentum, baryon number and color charge
conservation, is colorful {\it i.e.} there is a non-vanishing color
current in the system. The baryon chemical potentials of quarks and
of antiquarks and the scalar (colorless) chemical potential of
gluons are constrained as in a global equilibrium: 
$\bar \mu_b = - \mu_b$ and $\mu_g^0 = 0$. The local equilibrium
configuration resulting from the cancellation of collision terms,
which represent the most probable binary parton interactions,
is also colorful with the same color structure. The colorless
chemical potentials, however, are unconstrained. The global equilibrium
relations among them emerge when the subdominant processes are taken 
into account. It is conjectured that not only binary but even multi-parton
collisions comply with the finite color chemical potentials, thus 
suggesting that the color neutralization of the plasma occurs not due 
to the collisions but due to  dissipative collective phenomena. 
Proper identification of these processes and their quantitative 
description will be very important for understanding of the whole 
equilibration scenario of the quark-gluon plasma.

\acknowledgements

We are specially indebted to L.~Alvarez-Gaum\'e for very helpful
discussions. Useful conversations with A.~Pich, N.~Rius and V.~Semikoz
are also gratefully acknowledged. C.~M. thanks for the warm hospitality 
at Institute of Physics of \'Swi\c etokrzyska Academy in Kielce.
C.~M. was supported by the Generalitat Valenciana, under grant ctidia/2002/5.

\appendix


\section{Evaluation of traces}


We collect here some useful formulas of the traces computed both
in the fundamental and adjoint representation.

Due to the identity
\be \label{id-fund}
\tau^a_{ij}\tau^a_{kl} = {1 \over 2}\, \delta^{il}\delta^{jk}
- {1 \over 2 N_c}\, \delta^{ij}\delta^{kl} \;,
\ee
we have the relations of the traces in  the fundamental representation
\ba
\label{traces-fund1}
{\rm Tr} [\tau^a A \; \tau^a B] & = &
- {1 \over 2 N_c}\, {\rm Tr} [A B]
+ {1 \over 2}\, {\rm Tr} [ A ] \; {\rm Tr}[ B] \;, \\
{\rm Tr} [\tau^a A] \; {\rm Tr} [\tau^a B] & = &
{1 \over 2}\, {\rm Tr} [A B]
- {1 \over 2 N_c}\, {\rm Tr} [ A ] \; {\rm Tr}[ B] \;.
\ea

Furthermore, from Eq.~(\ref{id-fund}) one can deduce
\be \label{id-fund-secun}
\tau^a \tau^a =  \frac{N_c^2 -1}{2N_c} \;,   \qquad 
\tau^a \tau^b \tau^a= - \frac{1}{2 N_c} \tau^b
\ee

Taking into account
\be
\tau_a \tau_b = \frac{1}{2 N_c} \delta_{ab} + \frac 12 d_{abc} \tau_c
+  \frac i2 f_{abc} \tau_c
\ee
one evaluates traces of products of generators in the
fundamental representation. In particular, one finds
\ba
\label{3-tau}
{\rm Tr} [\tau^a \tau^b \tau^c] & = &  \frac{1}{4 } 
\Big( d^{abc} + i f^{abc} \Big) \;, \\
\label{4-tau}
{\rm Tr} [\tau^a \tau^b \tau^c \tau^d] & = & \frac{1}{4 N_c} \Big(
\delta^{ab} \delta^{cd} - \delta^{ac} \delta^{bd}
+ \delta^{ad} \delta^{bc} \Big) + 
\frac{1}{8} \Big(
d^{abr} d^{cdr} - d^{acr} d^{bdr}
+ d^{adr} d^{bcr} \Big) \nonumber
\\
& +& \frac{i}{8} \Big(
d^{abr} f^{cdr} - d^{acr} f^{bdr}
+ d^{adr} f^{bcr} \Big) 
\ea

For $N_c =2$ one also has
\be \label{5-tau}
{\rm Tr} [\tau^a \tau^b \tau^c \tau^d \tau^e]  =  \frac{i}{16} \Big(
\delta^{ae} f^{bcd} + \delta^{cd} f^{abe} + \delta^{bd} f^{aec} +
\delta^{bc} f^{ade} \Big)  
\ee

The identity analogous to (\ref{id-fund}) for the adjoint
representation of the SU(2) group is
\be \label{id-ad-2}
T^a_{bc}T^a_{de} =
 \delta^{be}\delta^{cd} - \delta^{bd}\delta^{ce} \;,
\ee
and we have the following relations
\ba
\label{traces-adjSU2}
{\rm Tr} [T^a A \; T^a B] & = &
{\rm Tr} [ A ] \; {\rm Tr}[ B] - {\rm Tr} [A B^T] \;, \\
{\rm Tr} [T^a A] \; {\rm Tr} [T^a B] & = &
{\rm Tr} [ A  B]  - {\rm Tr} [A B^T] \;.
\ea

Using the identity (\ref{id-ad-2}) one also finds
\begin{mathletters}
\label{tracesSU2}
\ba
{\rm Tr} [T^a T^b T^c] &=& i f^{abc} \;,
\\
{\rm Tr} [T^a T^b T^c T^d] &=& \delta^{ab} \delta^{cd} 
+ \delta^{ad} \delta^{bc} \;,
\\
{\rm Tr} [T^a T^b T^c T^d T^e ] &=& \delta^{ad} f^{ecb} 
- \delta^{cd} f^{eab} - \delta^{ab} f^{ecd} \;.
\ea
\end{mathletters}

In the adjoint representation of SU(3), we have the identity
\ba
T^a_{bc} T^a_{de} = - \frac 23 \left( \delta^{bd} \delta^{ce} -
\delta^{be} \delta^{cd} \right) -  \left( d_{bdf} d_{cef} -
d_{bef} d_{cdf} \right) \;,
\ea
which, in particular, allows one to compute the totally symmetric
trace of 4 generators as
\be
\label{4symmtr}
\frac 14 {\rm Tr} [\{T_a,T_b\}\{T_c,T_d\}] =
\frac 12 \left( 2\delta^{ab} \delta_{cd} + \delta^{ac} \delta_{bd} +
 \delta^{ad} \delta_{bc} \right) + \frac{3}{4} d_{abs} d_{cds} \ .
\ee



\end{document}